\documentclass[reprint, amsfonts, amssymb, amsmath, aps, prl, floatfix, twoside, superscriptaddress]{revtex4-1}
\usepackage{graphicx}
\usepackage{dcolumn}
\usepackage[english]{babel}
\usepackage[utf8]{inputenc}
\usepackage{amsthm}
\usepackage{mathtools}
\usepackage{physics}
\usepackage{xcolor,colortbl}
\usepackage{adjustbox}
\usepackage{placeins}
\usepackage[T1]{fontenc}
\usepackage{lipsum}
\usepackage{csquotes}
\usepackage{hyperref}
\usepackage{comment}
\usepackage[bb=boondox]{mathalfa}
\usepackage[normalem]{ulem}
\usepackage{diagbox}
\usepackage{slashbox}

\begin{document}
\title{Benchmarking the readout of a superconducting qubit for repeated measurements}
\author{S. Hazra}
\email{sumeru.hazra@yale.edu, wei.dai.wd279@yale.edu}
\thanks{These two authors contributed equally}
\author{W. Dai}
\email{sumeru.hazra@yale.edu, wei.dai.wd279@yale.edu}
\thanks{These two authors contributed equally}
\author{T. Connolly}
\author{P. D. Kurilovich}
\author{Z. Wang}
\author{L. Frunzio}
\author{M. H. Devoret}
\email{michel.devoret@yale.edu}
\affiliation{Department of Applied Physics, Yale University, New Haven, Connecticut 06520, USA\\
and Yale Quantum Institute, Yale University, New Haven, Connecticut 06520, USA}

\begin{abstract}
Readout of superconducting qubits faces a trade-off between measurement speed and unwanted back-action on the qubit caused by the readout drive, such as $T_1$ degradation and leakage out of the computational subspace.
The readout is typically benchmarked by integrating the readout signal and choosing a binary threshold to extract the ``readout fidelity''.
We show that readout fidelity may significantly overlook readout-induced leakage errors. Such errors are detrimental for applications that rely on continuously repeated measurements, e.g., quantum error correction.
We introduce a method to measure the readout-induced leakage rate by repeatedly executing a composite operation \textemdash a readout preceded by a randomized qubit-flip.
We apply this technique to characterize the  readout of 
a superconducting qubit, optimized for fidelity across four different readout durations.
Our technique highlights the importance of an independent leakage characterization by showing that the leakage rates vary from $0.12\%$ to $7.76\%$ across these readouts even though the fidelity exceeds $99.5\%$ in all four cases.
\end{abstract}
\maketitle

Fast and accurate single-shot qubit readout is crucial
for quantum computing experiments including measurement-based state preparation~\cite{msp_riste_2012}, entanglement generation~\cite{mbe_cabrillo_1999, mbe_lalumiere_2010, mbe_riste_2013}, and quantum error correction~\cite{ofek_qec, hu_2019_binomial, sivak_gkp, eth_qec_2022, google_qec, ni_qec}. 
The qubit readout in superconducting circuits is typically realized by 
probing a readout resonator dispersively coupled to a qubit~\cite{gambetta_2007}.
Ideally, increasing the power of the readout pulse in such a dispersive readout scheme should result in a larger signal-to-noise ratio (SNR) and consequently higher readout speed and fidelity. 
However, at higher readout power, the superconducting circuit housing the qubit undergoes spurious  transitions into non-computational ``leakage'' states~\cite{mist_sank_et_al, transmon_ionization, khezri_2023, unified_ionization}. The onset of deleterious transitions at higher power creates a tension between improving the SNR and containing the leakage.

Readout-induced leakage is particularly detrimental for applications that require continuously repeated measurements.
For example, in quantum error correction, efficient entropy removal from the quantum system is achieved by repeated readout and reset of the auxiliary qubits. 
A single leakage error leaves the qubit in a highly excited state and corrupts the reliability of future measurements. The presence of such correlated errors degrades the performance of error correction codes~\cite{Ghosh_2013,Fowler_2013}, especially when these errors spread into neighboring qubits through entangling gates~\cite{varbanov2020leakage}.
Thus, leakage errors pose a greater threat to quantum error correction~\cite{eth_qec_2022,miao_2023}, compared to the discrimination errors or Pauli errors. 

\begin{figure}[b!]
    \centering\includegraphics{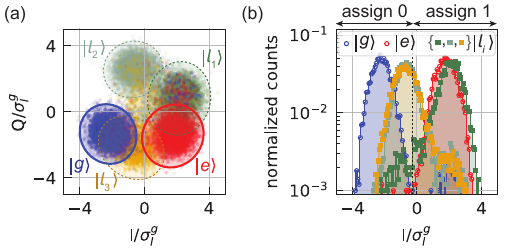}
    \caption{
    Overlap of the IQ signal distribution of the leakage states and the computational states.
    (a) Experimentally obtained readout signal in our device. The plot shows the IQ signals of $|g\rangle$ and $|e\rangle$  along with a few example non-computational states labeled $|l_i\rangle$~\cite{supp}. 
    The ellipses representing the 3$\sigma$ boundary are overlapping, indicating that discrimination of the non-computational states is challenging, even with a multi-threshold demarcation~\cite{mist_sank_et_al}. These prepared leakage states are shown  to illustrate how the readout signals of the leakage states can appear in the IQ plane, and may not be in general the leakage states limiting our readout.
    (b) Leakage state discrimination from a binary threshold (see dotted line) becomes even more difficult than in (a). Such a projection is economical from a signal processing perspective, and is unavoidable if the readout chain contains a phase-sensitive amplifier. 
    }
    \label{fig:concept}
\end{figure}

However, the metrics commonly used to characterize and optimize the readout performance may overlook leakage. As an example, the readout performance is often characterized by ``readout fidelity''~\cite{msp_riste_2012, jeffrey_2014, bultink_2016, walter_2017, touzard_cd, dassonneville_2020, sunada_2022} defined as $\mathcal{F}=\left[P(0|g)+P(1|e)\right]/2$, where $P(x|\psi)$ is the probability that the measurement outcome is $x$ given the qubit is prepared in $|\psi\rangle$. 
However, even a high-fidelity readout may overlook readout-induced leakage if it cannot distinguish from $|g\rangle$ or $|e\rangle$ the non-computational states into which the qubit leaks. In principle, the leakage error should be reflected in any metric quantifying the quantum non-demolition (QND) character \cite{Braginsky_qnd, caves_qnd, qnd_review} of the readout. A commonly used \textit{proxy} for the readout ``QNDness'' is the correlation of two successive readout outcomes~\cite{sunada_2022, spring_multiplexed_2024} $\mathcal{R} = [P(0|0, g)+P(1|1, e)]/2$, where $P(x|y, \psi)$ is the probability that the second readout outcome is $x$ given the qubit is prepared in $|\psi\rangle$ and the first readout outcome is $y$. Unfortunately, while this ``repeatability'' metric captures the bit-flip errors, it overlooks the leakage errors in the same manner as the fidelity metric. The relationship between readout fidelity, repeatability, and the QNDness in the presence of readout induced leakage is presented in Sec.~V of the supplemental material~\cite{supp}.

When the readout is optimized for speed and fidelity, even a full quadrature readout cannot effectively distinguish the leakage states.
To illustrate this point, in Fig.~\ref{fig:concept}, we show the overlapping readout signals of a few prepared non-computational states, together with $|g\rangle$ and $|e\rangle$. From the figure it is clear that the signal corresponding to the leakage states can be erroneously assigned to either ``g'' or ``e''. The situation is further complicated by the fact that the readout-induced leakage might involve several highly-excited states~\cite{unified_ionization}, leading to a smeared blob in the IQ plane~\cite{mist_sank_et_al}. In addition, the readout signal of leakage states may even appear in the middle of the $|g\rangle$ and $|e\rangle$ blobs (see $|l_3\rangle$ in Fig.~\ref{fig:concept}) when the dispersive shift of such a state is smaller than that of $|e\rangle$. Readout-induced leakage into these states cannot be distinguished, even with a detuned readout~\cite{chen_transmon}.
Moreover, the IQ data is usually projected along one quadrature and a binary threshold is used for state assignment~\cite{multiplexed_eth, msp_riste_2012, jeffrey_2014}. Such a projection worsens the overlap (for example, see the histograms for $|l_1\rangle$ and $|e\rangle$). 
The projection becomes inevitable if a phase-sensitive quantum-limited amplifier is used to improve the readout efficiency~\cite{dassonneville_2020, walter_2017, sunada_2022, touzard_cd}. In this case, the information on the other quadrature is erased, making it impossible to discriminate the leakage states from the IQ data. We are then led to the question of how to quantify the readout-induced leakage with a readout operation that itself cannot distinguish all the leakage states from $|g\rangle$ and $|e\rangle$.

In this Letter, we introduce a novel technique, ``readout-induced leakage benchmarking (RILB)'' to measure the leakage rate by repeating a composite operation\textemdash a readout preceded by a random qubit flip. 
We implement this technique to characterize the readout-induced leakage error on a Purcell-protected transmon~\cite{gambetta_2011, diniz_2013, roy_2017, dassonneville_2020, pfeiffer_2024}.
We optimize the readout pulses (frequency, envelope, and power) to achieve maximum fidelity for a set of different readout durations,
and show that readouts with nominally identical fidelities and repetabilities can have leakage rates differing by more than one order of magnitude.
These findings highlight that quantifying  the readout-induced leakage rate is a necessary step in optimizing the performance of quantum processors.

\begin{figure}[t!]
    \centering
    \includegraphics{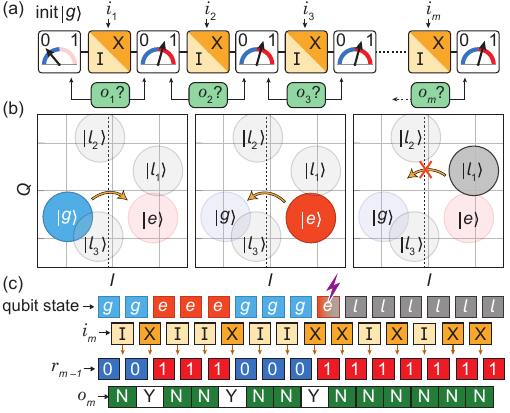}
    \caption{Principle of readout-induced leakage benchmarking (RILB). (a) Pulse sequence composed of interleaved readout operations and randomized bit-flips to measure the leakage rate. The symbols $i_m$ and $o_m$ represent the input and output bit-strings respectively. (b) The action of a $\pi$-pulse on successive readout outcomes. When the post readout qubit state is either $|g\rangle$ or $|e\rangle$, the $\pi$-pulse flips successive readout outcomes. When the qubit is exited to a leakage state, the $\pi$-pulse is ineffective and the readout outcome is either unchanged (e.g. $|l_1\rangle$) or random (e.g. $|l_2\rangle$). (c) A qubit-state simulation showing how a leakage event  (marked by the violet thunderbolt) affects the readout outcomes during RILB. The output bit string $o_m$ indicates whether the readout outcome flips (Y, standing for ``yes'') or does not flip (N, standing  for ``no'') from the previous one. In absence of any error, this bit string is perfectly correlated with the input ``X-or-I'' bit string $i_m$. If the qubit leaves the computational subspace, the readout outcomes are uncorrelated with the input operations for all subsequent rounds (strings of N's at the end of the last row).
    }
    \label{fig:schematic}
\end{figure}

\textit{Readout-induced leakage benchmarking.}\textemdash
Our proposed benchmarking technique interleaves repeated readout operations with randomized qubit flips (X), as depicted in Fig.~\ref{fig:schematic}(a).  
The qubit responds to the flip operation if and only if it is in the computational subspace and thus the outcomes of the successive readout operations evolve as shown in Fig.~\ref{fig:schematic}(c). 
A leakage error spoils the agreements for multiple rounds, as the qubit remains out of the computational subspace for several cycles.

To quantify the leakage rate, we perform the following steps. (1) Choose a sequence length that represents the total number of cycles, each consisting of a qubit operation and a readout. (2) Construct different randomizations for the input ``X-or-I'' bit-string $\{i_m\}$ and generate the corresponding interleaved pulse sequences, as illustrated in Fig.~\ref{fig:schematic}(a). (3) From the outcomes $r_m$ and $r_{m-1}$ of successive readouts, obtain the output bit-string $\{ o_m\}$ that detects ``flipped-or-not''. (4) Compute the bit-wise correlation $\mathcal{C}_m\in\{-1, +1\}$
for each experimental realization of the sequence (5) Perform multiple experimental realizations to compute the average bit-wise correlation $\bar{\mathcal{C}}_m$ for each random sequence. (6) Average over different randomizations to obtain the mean success probability $\langle \bar{\mathcal{C}} \rangle_m$ to identify the input operation.

To the leading order, the mean success probability will decay exponentially against the number of operations ($m$), given by the equation~\cite{wood_lrb_2018},
\begin{equation}
    \langle \bar{\mathcal{C}}\rangle_m=\frac{1}{2}\left(A+ B(1-L)^m\right),
    \label{eq:leakage_rate}
\end{equation}
where $L$ represents the average leakage probability per readout (leakage rate). 
Note that a Pauli error or a discrimination error only corrupts the correlation locally, and does not affect the decay constant. 
The effect of the discrimination error, state initialization error and Pauli error during the sequence are contained in $A$ and $B$ (see Sec.~VI of the supplemental material~\cite{supp} for details). Importantly, RILB technique leverages the correlated nature of the leakage errors in continuously repeated measurements.
As a result, RILB efficiently measures the readout-induced leakage rate, even if the leakage probability per readout is much smaller compared to other error probabilities.

\begin{figure}[t]
\includegraphics[width = 0.48\textwidth]{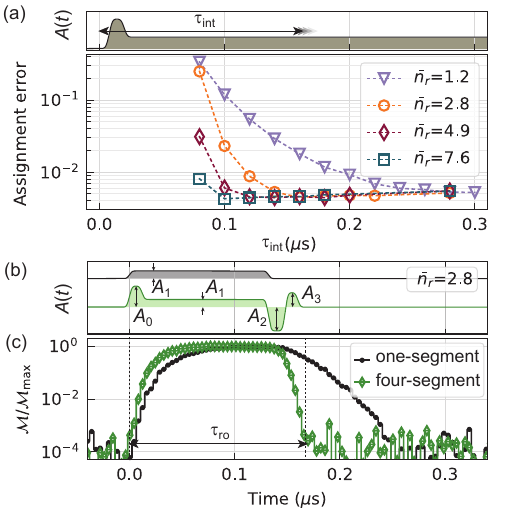}
\caption
[Readout assignment error vs integration time]
{Readout pulse calibration. (a) Two-step readout pulse with an initial high power segment and (below)
the readout assignment error plotted against integration time $\tau_{\rm{int}}$ for varying readout powers (represented by steady-state intra-resonator photon number $\bar{n}_r$). 
(b) Envelopes of the one-segment and four-segment readout pulses, both calibrated for a readout with $\bar{n}_r=2.8$ corresponding to the amplitude segment $A_1$ for each pulse. 
Note that, the resonator does not reach the steady state for short ($\tau\sim\kappa_r^{-1}$) pulses and $\bar{n}_r$ is only used as a scale for the drive power.
(c) Normalized qubit measurement rates as a function of time. 
The readout duration, $\tau_{\rm{ro}}$, is defined as the interval when the readout resonator is populated.
}
\label{fig:2} 
\end{figure}

\textit{Readout pulse calibration}.\textemdash 
We apply the RILB technique to characterize the leakage caused by the dispersive readout~\cite{wallraff_disp_2004} of an intrinsically Purcell-protected transmon embedded in a multimodal circuit~\cite{supp}. However, the scope of this technique is not limited to dispersive readout and can be extended to continuously repeated measurements of any qubit. The detailed description of our experimental device and the measurement setup are respectively presented in Sec.~II and III of the supplemental materials~\cite{supp}. In our implementation, the qubit state dependent total dispersive shift of the readout resonator is 
$6.4$ MHz and the readout resonator has an external coupling rate of $\kappa_r/2\pi = 11.6$ MHz.
To determine the choice of readout duration and power, we apply a two-step readout pulse~\cite{walter_2017}, depicted in Fig.~\ref{fig:2}(a). Unlike a one-segment ``boxcar'' pulse, this technique ramps up the resonator faster than its natural ring-up time ($\kappa_r^{-1}$), improving the
SNR at short duration ($\tau\sim \kappa_r^{-1}$)~\cite{supp}.
The assignment error of the binary readout, $\left[P(0|e)+P(1|g)\right]/2$, is evaluated from $100\,000$ shots. 
In Fig.~\ref{fig:2}(a), we plot the assignment error against varying integration times $\tau_{\rm{int}}$ and steady-state photon number $\bar{n}_r$ in the resonator. 
This plot instructs us on the optimal readout duration for a given readout power.

It is worth noting that for repeated measurements, the integration time does not correspond to the ``readout duration'' of practical interest. 
In such applications, a relevant definition of readout duration should include the time during which the qubit is \textit{unavailable for other operations}.
After the ``boxcar'' readout drive is turned off, the tail of the readout signal persists and \textit{continues to dephase the qubit} for a timescale $\tau \sim 10/\kappa_r$, preventing any high-fidelity qubit gates.  This is shown by plotting the measurement rate, $\mathcal{M}(t) \propto |\langle\alpha_g(t)-\alpha_e(t)\rangle|^2$ (black) in Fig.~\ref{fig:2}(c),
Where $\alpha_{g,e}(t)$ are the instantaneous complex amplitudes of readout resonator corresponding to the two states of the qubit. Therefore, the readout duration refers to the interval when the readout resonator is populated \textemdash from the beginning of the readout pulse to the instant beyond which the readout-photon-induced dephasing rate is inconsequential compared to the bare dephasing rate of the qubit. 

We experimentally optimize a four-segment pulse~\cite{jeffrey_2014, bultink_2016, mcclure_2016} to minimize the readout duration for a given SNR at each readout power, as shown in green in Fig.~\ref{fig:2}(b).
The last two segments of this composite pulse empty the readout resonator unconditionally (regardless of the qubit state: $|g\rangle$ or $|e\rangle$), as illustrated by green markers in  Fig.~\ref{fig:2}(c). See Sec.~IV of the supplemental materials~\cite{supp} for details.
The SNAIL parametric amplifier (SPA)~\cite{spa_opt} in our experiment fails to respond as fast as the outgoing readout wave-packet, due to its limited $3$ dB bandwidth of $8.3$ MHz at the operating point, causing a slow-down of the wave-packet.
Thus to capture the true resonator dynamics, the data in Fig.~\ref{fig:2}(c) is acquired with the SPA turned off (averaged over $100\,000$ shots). 
In a single-shot readout with the SPA on, we integrate for $140$~ns longer than the duration of the readout pulse itself to completely acquire the wave-packet leaving the SPA.
But this does not affect the readout duration defined above, as we confirm the restoration of qubit operations immediately after the readout pulse ends.

\begin{figure}[ht!]
\includegraphics[width = 0.45\textwidth]{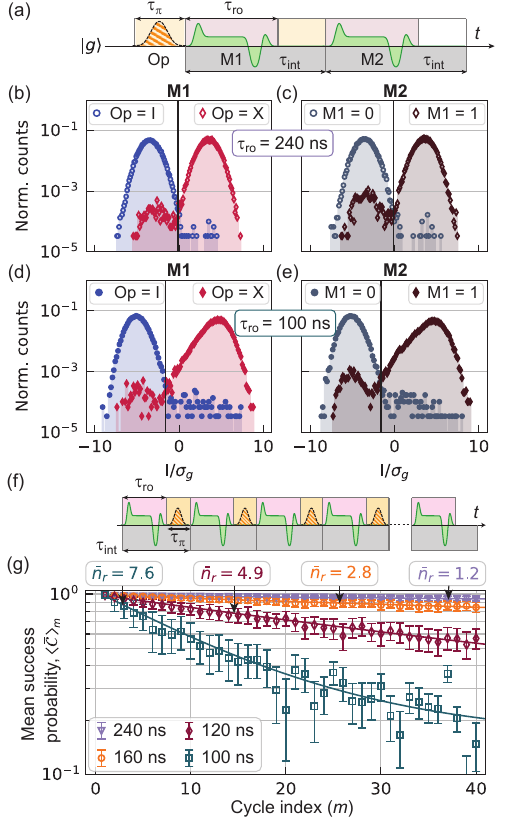}
\caption
{Characterization of readout fidelity, repeatability, and readout-induced leakage. (a) Pulse sequence to measure fidelity and repeatability. (b-e) Histograms of the integrated signal corresponding to the first ($\rm{M1}$) and the second ($\rm{M2}$) readout operations respectively for readout durations of (b-c)  $\tau_{\rm{ro}} = 240$ ns ($\bar{n}_r = 1.2$) and (d-e) $\tau_{\rm{ro}} = 100$ ns ($\bar{n}_r = 7.6$), constructed from $30\,000$ shots. The readout duration is determined for each readout power to maintain the same SNR.
The vertical lines represent the optimal demarcation thresholds in these settings. The readout-induced leakage at high-powers distorts the `$e$'-histogram from a Gaussian shape. This effect does not reflect in the readout outcome due to the choice of a \textit{binary-threshold}, labeling them as the `$e$'-state.
(f) Pulse sequence of the RILB experiment. (g) The measured mean success probability $\langle\bar{\mathcal{C}}\rangle_m$ is plotted as a function of cycle index $m$ for  four different readout operations ($100$ ns, $120$ ns, $160$ ns and $240$ ns), each optimized for fidelity. The corresponding (steady state) readout photon numbers are also annotated above the plot. Solid curves correspond to exponential fits.
}
\label{fig:characterization} 
\end{figure}

\textit{Single-shot readout performance.}\textemdash
Increasing the readout power speeds up the readout operation but causes an increased leakage rate. To compare the readout performance, we optimize the readout pulse across four different readout durations and first characterize them by measuring the fidelity and repeatability~\cite{sunada_2022, swiadek_2023}. We initialize the qubit with pre-selection and prepare it in $|g\rangle$ or $|e\rangle$, and then apply two successive readout pulses, as shown in Fig.~\ref{fig:characterization}(a).
The binary readout fidelity $\mathcal{F}$ and the readout repeatability $\mathcal{R}$ are then calculated from the integrated readout histograms (see Fig.~\ref{fig:characterization}(b-e)) using the equations mentioned in the introduction.
The readout duration, fidelity and repeatability for each of these readout settings are listed in Table~\ref{table:readout_performance}. 
The distortion of the `$e$' histogram from a Gaussian shape in Fig.~\ref{fig:characterization}(d-e) suggests that there is significant readout-induced leakage to higher states. Such distortions however do not affect the fidelity or repeatability of the binary readout.

\textit{Experimental leakage characterization.}\textemdash
To estimate the leakage rate, we implement the RILB technique with the same four readout pulses. 
We use a $100$~ns $\pi$-pulse so that the spectral content of the pulse is smaller than the anharmonicity of the qubit and the leakage rate from the bit-flip operation itself can be neglected. The $\pi$-pulse is played during the padding time $\tau_{\rm{pad}} = (\tau_{\rm{int}}-\tau_{\rm{ro}}) = 140$~ns between two readout pulses, while the previous readout wave-packet is still being acquired.
We choose a sequence consisting of $40$ readout operations with $98$ different randomizations 
and evaluate the mean success probability, $\langle \bar{\mathcal{C}}\rangle_m $ 
for each randomization. 
$\langle \bar{\mathcal{C}}\rangle_m $ decays exponentially with $m$ due to the accumulation of leakage errors, as shown in Fig.~\ref{fig:characterization}(f). 
We fit the data to the exponential decay model shown in Eq.~\ref{eq:leakage_rate} and list the average leakage rate $L$ for each of the four optimized readout pulses in Table~\ref{table:readout_performance}. 
Note that the readout-induced leakage per readout operation rises from $0.12\%$ to $7.76\%$ as we increase the readout speed. This degradation is not reflected in readout fidelity or repeatability of these four readout pulses.
Our result thus highlights that, in addition to readout duration, fidelity, and repeatability, one must characterize the readout-induced leakage. 

\begin{table}[t]
\centering
\begin{tabular}{|l|c|c|c|c|}
    \hline 
    RO duration, $\tau_{\rm{ro}}$ (ns) & $240$ & $160$ & $120$ & $100$\\
    \hline
    Photon number, $\bar{n}_r$ & $1.2$ & $2.8$ & $4.9$ & $7.6$\\
    Fidelity, $\mathcal{F}$ (\%) & $99.63$ & $99.63$ & $99.61$ & $99.54$\\
    Repeatability, $\mathcal{R}$ (\%)  & $99.12$ & $99.11$ & $99.04$ & $99.01$\\
    Leakage per RO, $L$(\%) & $0.12$ & $0.48$ & $2.14$ & $7.76$\\
    \hline
\end{tabular}
\caption{Binary readout (RO) metrics\textemdash Readout duration, steady-state photon number, fidelity, repeatability, and average leakage rate (measured by RILB) for four different readout pulses, each optimized for fidelity. 
Faster readouts with a higher powers result in significantly increased leakage rates.
Such a degradation is not captured by fidelity or repeatability. }
\label{table:readout_performance}
\end{table}

\textit{Conclusions.}\textemdash Basing ourselves on a state-of-the art dispersive readout in terms of speed, fidelity, and repeatability, we issue a warning that the readout fidelity and/or repeatability of a superconducting qubit should not be the sole optimization guideline. 
We introduce a readout-induced leakage characterization technique that requires only binary-thresholded readouts to quantify the leakage rate. Readout-induced leakage should be independently quantified and minimized while tuning up a readout pulse for repeated measurements. 
Future work will involve a systematic investigation of the mechanisms~\cite{unified_ionization} for such readout-induced leakage and further improvement of qubit readout at hardware level. 

\begin{acknowledgments}
The authors acknowledge  R. Cortiñas, V.R. Joshi, A. Koottandavida, W. Kalfus, A. Miano, P.D. Parakh, J. Venkatraman, D. Weiss, and X. Xiao for illuminating discussions.
This research was sponsored by the Army Research Office (ARO) under grant nos. W911NF-23-1-0051, by the Air Force Office of Scientific Research (AFOSR) under grant FA9550-19-1-0399 and by the U.S. Department of Energy (DoE), Office of Science, National Quantum Information Science Research Centers, Co-design Center for Quantum Advantage (C2QA) under contract number DE-SC0012704. The views and conclusions contained in this document are those of the authors and should not be interpreted as representing the official policies, either expressed or implied, of the ARO, AFOSR, DoE or the US Government. The US Government is authorized to reproduce and distribute reprints for Government purposes notwithstanding any copyright notation herein. Fabrication facilities use was supported by the Yale Institute for Nanoscience and Quantum Engineering (YINQE) and the Yale SEAS Cleanroom.
L.F. is a founder and shareholder of Quantum Circuits Inc. (QCI).
\end{acknowledgments}


\begin{thebibliography}{41}%
\makeatletter
\providecommand \@ifxundefined [1]{%
 \@ifx{#1\undefined}
}%
\providecommand \@ifnum [1]{%
 \ifnum #1\expandafter \@firstoftwo
 \else \expandafter \@secondoftwo
 \fi
}%
\providecommand \@ifx [1]{%
 \ifx #1\expandafter \@firstoftwo
 \else \expandafter \@secondoftwo
 \fi
}%
\providecommand \natexlab [1]{#1}%
\providecommand \enquote  [1]{``#1''}%
\providecommand \bibnamefont  [1]{#1}%
\providecommand \bibfnamefont [1]{#1}%
\providecommand \citenamefont [1]{#1}%
\providecommand \href@noop [0]{\@secondoftwo}%
\providecommand \href [0]{\begingroup \@sanitize@url \@href}%
\providecommand \@href[1]{\@@startlink{#1}\@@href}%
\providecommand \@@href[1]{\endgroup#1\@@endlink}%
\providecommand \@sanitize@url [0]{\catcode `\\12\catcode `\$12\catcode `\&12\catcode `\#12\catcode `\^12\catcode `\_12\catcode `\%12\relax}%
\providecommand \@@startlink[1]{}%
\providecommand \@@endlink[0]{}%
\providecommand \url  [0]{\begingroup\@sanitize@url \@url }%
\providecommand \@url [1]{\endgroup\@href {#1}{\urlprefix }}%
\providecommand \urlprefix  [0]{URL }%
\providecommand \Eprint [0]{\href }%
\providecommand \doibase [0]{http://dx.doi.org/}%
\providecommand \selectlanguage [0]{\@gobble}%
\providecommand \bibinfo  [0]{\@secondoftwo}%
\providecommand \bibfield  [0]{\@secondoftwo}%
\providecommand \translation [1]{[#1]}%
\providecommand \BibitemOpen [0]{}%
\providecommand \bibitemStop [0]{}%
\providecommand \bibitemNoStop [0]{.\EOS\space}%
\providecommand \EOS [0]{\spacefactor3000\relax}%
\providecommand \BibitemShut  [1]{\csname bibitem#1\endcsname}%
\let\auto@bib@innerbib\@empty
\bibitem [{\citenamefont {Rist\`e}\ \emph {et~al.}(2012)\citenamefont {Rist\`e}, \citenamefont {van Leeuwen}, \citenamefont {Ku}, \citenamefont {Lehnert},\ and\ \citenamefont {DiCarlo}}]{msp_riste_2012}%
  \BibitemOpen
  \bibfield  {author} {\bibinfo {author} {\bibfnamefont {D.}~\bibnamefont {Rist\`e}}, \bibinfo {author} {\bibfnamefont {J.~G.}\ \bibnamefont {van Leeuwen}}, \bibinfo {author} {\bibfnamefont {H.-S.}\ \bibnamefont {Ku}}, \bibinfo {author} {\bibfnamefont {K.~W.}\ \bibnamefont {Lehnert}}, \ and\ \bibinfo {author} {\bibfnamefont {L.}~\bibnamefont {DiCarlo}},\ }\href {\doibase 10.1103/PhysRevLett.109.050507} {\bibfield  {journal} {\bibinfo  {journal} {Phys. Rev. Lett.}\ }\textbf {\bibinfo {volume} {109}},\ \bibinfo {pages} {050507} (\bibinfo {year} {2012})}\BibitemShut {NoStop}%
\bibitem [{\citenamefont {Cabrillo}\ \emph {et~al.}(1999)\citenamefont {Cabrillo}, \citenamefont {Cirac}, \citenamefont {Garc\'{\i}a-Fern\'andez},\ and\ \citenamefont {Zoller}}]{mbe_cabrillo_1999}%
  \BibitemOpen
  \bibfield  {author} {\bibinfo {author} {\bibfnamefont {C.}~\bibnamefont {Cabrillo}}, \bibinfo {author} {\bibfnamefont {J.~I.}\ \bibnamefont {Cirac}}, \bibinfo {author} {\bibfnamefont {P.}~\bibnamefont {Garc\'{\i}a-Fern\'andez}}, \ and\ \bibinfo {author} {\bibfnamefont {P.}~\bibnamefont {Zoller}},\ }\href {\doibase 10.1103/PhysRevA.59.1025} {\bibfield  {journal} {\bibinfo  {journal} {Phys. Rev. A}\ }\textbf {\bibinfo {volume} {59}},\ \bibinfo {pages} {1025} (\bibinfo {year} {1999})}\BibitemShut {NoStop}%
\bibitem [{\citenamefont {Lalumi\`ere}\ \emph {et~al.}(2010)\citenamefont {Lalumi\`ere}, \citenamefont {Gambetta},\ and\ \citenamefont {Blais}}]{mbe_lalumiere_2010}%
  \BibitemOpen
  \bibfield  {author} {\bibinfo {author} {\bibfnamefont {K.}~\bibnamefont {Lalumi\`ere}}, \bibinfo {author} {\bibfnamefont {J.~M.}\ \bibnamefont {Gambetta}}, \ and\ \bibinfo {author} {\bibfnamefont {A.}~\bibnamefont {Blais}},\ }\href {\doibase 10.1103/PhysRevA.81.040301} {\bibfield  {journal} {\bibinfo  {journal} {Phys. Rev. A}\ }\textbf {\bibinfo {volume} {81}},\ \bibinfo {pages} {040301} (\bibinfo {year} {2010})}\BibitemShut {NoStop}%
\bibitem [{\citenamefont {Riste}\ \emph {et~al.}(2013)\citenamefont {Riste}, \citenamefont {Dukalski}, \citenamefont {Watson}, \citenamefont {De~Lange}, \citenamefont {Tiggelman}, \citenamefont {Blanter}, \citenamefont {Lehnert}, \citenamefont {Schouten},\ and\ \citenamefont {DiCarlo}}]{mbe_riste_2013}%
  \BibitemOpen
  \bibfield  {author} {\bibinfo {author} {\bibfnamefont {D.}~\bibnamefont {Riste}}, \bibinfo {author} {\bibfnamefont {M.}~\bibnamefont {Dukalski}}, \bibinfo {author} {\bibfnamefont {C.}~\bibnamefont {Watson}}, \bibinfo {author} {\bibfnamefont {G.}~\bibnamefont {De~Lange}}, \bibinfo {author} {\bibfnamefont {M.}~\bibnamefont {Tiggelman}}, \bibinfo {author} {\bibfnamefont {Y.~M.}\ \bibnamefont {Blanter}}, \bibinfo {author} {\bibfnamefont {K.~W.}\ \bibnamefont {Lehnert}}, \bibinfo {author} {\bibfnamefont {R.}~\bibnamefont {Schouten}}, \ and\ \bibinfo {author} {\bibfnamefont {L.}~\bibnamefont {DiCarlo}},\ }\href@noop {} {\bibfield  {journal} {\bibinfo  {journal} {Nature}\ }\textbf {\bibinfo {volume} {502}},\ \bibinfo {pages} {350} (\bibinfo {year} {2013})}\BibitemShut {NoStop}%
\bibitem [{\citenamefont {Ofek}\ \emph {et~al.}(2016)\citenamefont {Ofek}, \citenamefont {Petrenko}, \citenamefont {Heeres}, \citenamefont {Reinhold}, \citenamefont {Leghtas}, \citenamefont {Vlastakis}, \citenamefont {Liu}, \citenamefont {Frunzio}, \citenamefont {Girvin}, \citenamefont {Jiang} \emph {et~al.}}]{ofek_qec}%
  \BibitemOpen
  \bibfield  {author} {\bibinfo {author} {\bibfnamefont {N.}~\bibnamefont {Ofek}}, \bibinfo {author} {\bibfnamefont {A.}~\bibnamefont {Petrenko}}, \bibinfo {author} {\bibfnamefont {R.}~\bibnamefont {Heeres}}, \bibinfo {author} {\bibfnamefont {P.}~\bibnamefont {Reinhold}}, \bibinfo {author} {\bibfnamefont {Z.}~\bibnamefont {Leghtas}}, \bibinfo {author} {\bibfnamefont {B.}~\bibnamefont {Vlastakis}}, \bibinfo {author} {\bibfnamefont {Y.}~\bibnamefont {Liu}}, \bibinfo {author} {\bibfnamefont {L.}~\bibnamefont {Frunzio}}, \bibinfo {author} {\bibfnamefont {S.~M.}\ \bibnamefont {Girvin}}, \bibinfo {author} {\bibfnamefont {L.}~\bibnamefont {Jiang}},  \emph {et~al.},\ }\href@noop {} {\bibfield  {journal} {\bibinfo  {journal} {Nature}\ }\textbf {\bibinfo {volume} {536}},\ \bibinfo {pages} {441} (\bibinfo {year} {2016})}\BibitemShut {NoStop}%
\bibitem [{\citenamefont {Hu}\ \emph {et~al.}(2019)\citenamefont {Hu}, \citenamefont {Ma}, \citenamefont {Cai}, \citenamefont {Mu}, \citenamefont {Xu}, \citenamefont {Wang}, \citenamefont {Wu}, \citenamefont {Wang}, \citenamefont {Song}, \citenamefont {Zou} \emph {et~al.}}]{hu_2019_binomial}%
  \BibitemOpen
  \bibfield  {author} {\bibinfo {author} {\bibfnamefont {L.}~\bibnamefont {Hu}}, \bibinfo {author} {\bibfnamefont {Y.}~\bibnamefont {Ma}}, \bibinfo {author} {\bibfnamefont {W.}~\bibnamefont {Cai}}, \bibinfo {author} {\bibfnamefont {X.}~\bibnamefont {Mu}}, \bibinfo {author} {\bibfnamefont {Y.}~\bibnamefont {Xu}}, \bibinfo {author} {\bibfnamefont {W.}~\bibnamefont {Wang}}, \bibinfo {author} {\bibfnamefont {Y.}~\bibnamefont {Wu}}, \bibinfo {author} {\bibfnamefont {H.}~\bibnamefont {Wang}}, \bibinfo {author} {\bibfnamefont {Y.}~\bibnamefont {Song}}, \bibinfo {author} {\bibfnamefont {C.-L.}\ \bibnamefont {Zou}},  \emph {et~al.},\ }\href@noop {} {\bibfield  {journal} {\bibinfo  {journal} {Nature Physics}\ }\textbf {\bibinfo {volume} {15}},\ \bibinfo {pages} {503} (\bibinfo {year} {2019})}\BibitemShut {NoStop}%
\bibitem [{\citenamefont {Sivak}\ \emph {et~al.}(2023)\citenamefont {Sivak}, \citenamefont {Eickbusch}, \citenamefont {Royer}, \citenamefont {Singh}, \citenamefont {Tsioutsios}, \citenamefont {Ganjam}, \citenamefont {Miano}, \citenamefont {Brock}, \citenamefont {Ding}, \citenamefont {Frunzio} \emph {et~al.}}]{sivak_gkp}%
  \BibitemOpen
  \bibfield  {author} {\bibinfo {author} {\bibfnamefont {V.}~\bibnamefont {Sivak}}, \bibinfo {author} {\bibfnamefont {A.}~\bibnamefont {Eickbusch}}, \bibinfo {author} {\bibfnamefont {B.}~\bibnamefont {Royer}}, \bibinfo {author} {\bibfnamefont {S.}~\bibnamefont {Singh}}, \bibinfo {author} {\bibfnamefont {I.}~\bibnamefont {Tsioutsios}}, \bibinfo {author} {\bibfnamefont {S.}~\bibnamefont {Ganjam}}, \bibinfo {author} {\bibfnamefont {A.}~\bibnamefont {Miano}}, \bibinfo {author} {\bibfnamefont {B.}~\bibnamefont {Brock}}, \bibinfo {author} {\bibfnamefont {A.}~\bibnamefont {Ding}}, \bibinfo {author} {\bibfnamefont {L.}~\bibnamefont {Frunzio}},  \emph {et~al.},\ }\href@noop {} {\bibfield  {journal} {\bibinfo  {journal} {Nature}\ }\textbf {\bibinfo {volume} {616}},\ \bibinfo {pages} {50} (\bibinfo {year} {2023})}\BibitemShut {NoStop}%
\bibitem [{\citenamefont {Krinner}\ \emph {et~al.}(2022)\citenamefont {Krinner}, \citenamefont {Lacroix}, \citenamefont {Remm}, \citenamefont {Di~Paolo}, \citenamefont {Genois}, \citenamefont {Leroux}, \citenamefont {Hellings}, \citenamefont {Lazar}, \citenamefont {Swiadek}, \citenamefont {Herrmann} \emph {et~al.}}]{eth_qec_2022}%
  \BibitemOpen
  \bibfield  {author} {\bibinfo {author} {\bibfnamefont {S.}~\bibnamefont {Krinner}}, \bibinfo {author} {\bibfnamefont {N.}~\bibnamefont {Lacroix}}, \bibinfo {author} {\bibfnamefont {A.}~\bibnamefont {Remm}}, \bibinfo {author} {\bibfnamefont {A.}~\bibnamefont {Di~Paolo}}, \bibinfo {author} {\bibfnamefont {E.}~\bibnamefont {Genois}}, \bibinfo {author} {\bibfnamefont {C.}~\bibnamefont {Leroux}}, \bibinfo {author} {\bibfnamefont {C.}~\bibnamefont {Hellings}}, \bibinfo {author} {\bibfnamefont {S.}~\bibnamefont {Lazar}}, \bibinfo {author} {\bibfnamefont {F.}~\bibnamefont {Swiadek}}, \bibinfo {author} {\bibfnamefont {J.}~\bibnamefont {Herrmann}},  \emph {et~al.},\ }\href@noop {} {\bibfield  {journal} {\bibinfo  {journal} {Nature}\ }\textbf {\bibinfo {volume} {605}},\ \bibinfo {pages} {669} (\bibinfo {year} {2022})}\BibitemShut {NoStop}%
\bibitem [{\citenamefont {AI}(2023)}]{google_qec}%
  \BibitemOpen
  \bibfield  {author} {\bibinfo {author} {\bibfnamefont {G.~Q.}\ \bibnamefont {AI}},\ }\href@noop {} {\bibfield  {journal} {\bibinfo  {journal} {Nature}\ }\textbf {\bibinfo {volume} {614}},\ \bibinfo {pages} {676} (\bibinfo {year} {2023})}\BibitemShut {NoStop}%
\bibitem [{\citenamefont {Ni}\ \emph {et~al.}(2023)\citenamefont {Ni}, \citenamefont {Li}, \citenamefont {Deng}, \citenamefont {Cai}, \citenamefont {Zhang}, \citenamefont {Wang}, \citenamefont {Yang}, \citenamefont {Yu}, \citenamefont {Yan}, \citenamefont {Liu} \emph {et~al.}}]{ni_qec}%
  \BibitemOpen
  \bibfield  {author} {\bibinfo {author} {\bibfnamefont {Z.}~\bibnamefont {Ni}}, \bibinfo {author} {\bibfnamefont {S.}~\bibnamefont {Li}}, \bibinfo {author} {\bibfnamefont {X.}~\bibnamefont {Deng}}, \bibinfo {author} {\bibfnamefont {Y.}~\bibnamefont {Cai}}, \bibinfo {author} {\bibfnamefont {L.}~\bibnamefont {Zhang}}, \bibinfo {author} {\bibfnamefont {W.}~\bibnamefont {Wang}}, \bibinfo {author} {\bibfnamefont {Z.-B.}\ \bibnamefont {Yang}}, \bibinfo {author} {\bibfnamefont {H.}~\bibnamefont {Yu}}, \bibinfo {author} {\bibfnamefont {F.}~\bibnamefont {Yan}}, \bibinfo {author} {\bibfnamefont {S.}~\bibnamefont {Liu}},  \emph {et~al.},\ }\href@noop {} {\bibfield  {journal} {\bibinfo  {journal} {Nature}\ }\textbf {\bibinfo {volume} {616}},\ \bibinfo {pages} {56} (\bibinfo {year} {2023})}\BibitemShut {NoStop}%
\bibitem [{\citenamefont {Gambetta}\ \emph {et~al.}(2007)\citenamefont {Gambetta}, \citenamefont {Braff}, \citenamefont {Wallraff}, \citenamefont {Girvin},\ and\ \citenamefont {Schoelkopf}}]{gambetta_2007}%
  \BibitemOpen
  \bibfield  {author} {\bibinfo {author} {\bibfnamefont {J.}~\bibnamefont {Gambetta}}, \bibinfo {author} {\bibfnamefont {W.~A.}\ \bibnamefont {Braff}}, \bibinfo {author} {\bibfnamefont {A.}~\bibnamefont {Wallraff}}, \bibinfo {author} {\bibfnamefont {S.~M.}\ \bibnamefont {Girvin}}, \ and\ \bibinfo {author} {\bibfnamefont {R.~J.}\ \bibnamefont {Schoelkopf}},\ }\href {\doibase 10.1103/PhysRevA.76.012325} {\bibfield  {journal} {\bibinfo  {journal} {Phys. Rev. A}\ }\textbf {\bibinfo {volume} {76}},\ \bibinfo {pages} {012325} (\bibinfo {year} {2007})}\BibitemShut {NoStop}%
\bibitem [{\citenamefont {Sank}\ \emph {et~al.}(2016)\citenamefont {Sank}, \citenamefont {Chen}, \citenamefont {Khezri}, \citenamefont {Kelly}, \citenamefont {Barends}, \citenamefont {Campbell}, \citenamefont {Chen}, \citenamefont {Chiaro}, \citenamefont {Dunsworth}, \citenamefont {Fowler}, \citenamefont {Jeffrey}, \citenamefont {Lucero} \emph {et~al.}}]{mist_sank_et_al}%
  \BibitemOpen
  \bibfield  {author} {\bibinfo {author} {\bibfnamefont {D.}~\bibnamefont {Sank}}, \bibinfo {author} {\bibfnamefont {Z.}~\bibnamefont {Chen}}, \bibinfo {author} {\bibfnamefont {M.}~\bibnamefont {Khezri}}, \bibinfo {author} {\bibfnamefont {J.}~\bibnamefont {Kelly}}, \bibinfo {author} {\bibfnamefont {R.}~\bibnamefont {Barends}}, \bibinfo {author} {\bibfnamefont {B.}~\bibnamefont {Campbell}}, \bibinfo {author} {\bibfnamefont {Y.}~\bibnamefont {Chen}}, \bibinfo {author} {\bibfnamefont {B.}~\bibnamefont {Chiaro}}, \bibinfo {author} {\bibfnamefont {A.}~\bibnamefont {Dunsworth}}, \bibinfo {author} {\bibfnamefont {A.}~\bibnamefont {Fowler}}, \bibinfo {author} {\bibfnamefont {E.}~\bibnamefont {Jeffrey}}, \bibinfo {author} {\bibfnamefont {E.}~\bibnamefont {Lucero}},  \emph {et~al.},\ }\href {\doibase 10.1103/PhysRevLett.117.190503} {\bibfield  {journal} {\bibinfo  {journal} {Phys. Rev. Lett.}\ }\textbf {\bibinfo {volume} {117}},\ \bibinfo {pages} {190503} (\bibinfo {year} {2016})}\BibitemShut {NoStop}%
\bibitem [{\citenamefont {Shillito}\ \emph {et~al.}(2022)\citenamefont {Shillito}, \citenamefont {Petrescu}, \citenamefont {Cohen}, \citenamefont {Beall}, \citenamefont {Hauru}, \citenamefont {Ganahl}, \citenamefont {Lewis}, \citenamefont {Vidal},\ and\ \citenamefont {Blais}}]{transmon_ionization}%
  \BibitemOpen
  \bibfield  {author} {\bibinfo {author} {\bibfnamefont {R.}~\bibnamefont {Shillito}}, \bibinfo {author} {\bibfnamefont {A.}~\bibnamefont {Petrescu}}, \bibinfo {author} {\bibfnamefont {J.}~\bibnamefont {Cohen}}, \bibinfo {author} {\bibfnamefont {J.}~\bibnamefont {Beall}}, \bibinfo {author} {\bibfnamefont {M.}~\bibnamefont {Hauru}}, \bibinfo {author} {\bibfnamefont {M.}~\bibnamefont {Ganahl}}, \bibinfo {author} {\bibfnamefont {A.~G.}\ \bibnamefont {Lewis}}, \bibinfo {author} {\bibfnamefont {G.}~\bibnamefont {Vidal}}, \ and\ \bibinfo {author} {\bibfnamefont {A.}~\bibnamefont {Blais}},\ }\href {\doibase 10.1103/PhysRevApplied.18.034031} {\bibfield  {journal} {\bibinfo  {journal} {Phys. Rev. Appl.}\ }\textbf {\bibinfo {volume} {18}},\ \bibinfo {pages} {034031} (\bibinfo {year} {2022})}\BibitemShut {NoStop}%
\bibitem [{\citenamefont {Khezri}\ \emph {et~al.}(2023)\citenamefont {Khezri}, \citenamefont {Opremcak}, \citenamefont {Chen}, \citenamefont {Miao}, \citenamefont {McEwen}, \citenamefont {Bengtsson}, \citenamefont {White}, \citenamefont {Naaman}, \citenamefont {Sank}, \citenamefont {Korotkov}, \citenamefont {Chen},\ and\ \citenamefont {Smelyanskiy}}]{khezri_2023}%
  \BibitemOpen
  \bibfield  {author} {\bibinfo {author} {\bibfnamefont {M.}~\bibnamefont {Khezri}}, \bibinfo {author} {\bibfnamefont {A.}~\bibnamefont {Opremcak}}, \bibinfo {author} {\bibfnamefont {Z.}~\bibnamefont {Chen}}, \bibinfo {author} {\bibfnamefont {K.~C.}\ \bibnamefont {Miao}}, \bibinfo {author} {\bibfnamefont {M.}~\bibnamefont {McEwen}}, \bibinfo {author} {\bibfnamefont {A.}~\bibnamefont {Bengtsson}}, \bibinfo {author} {\bibfnamefont {T.}~\bibnamefont {White}}, \bibinfo {author} {\bibfnamefont {O.}~\bibnamefont {Naaman}}, \bibinfo {author} {\bibfnamefont {D.}~\bibnamefont {Sank}}, \bibinfo {author} {\bibfnamefont {A.~N.}\ \bibnamefont {Korotkov}}, \bibinfo {author} {\bibfnamefont {Y.}~\bibnamefont {Chen}}, \ and\ \bibinfo {author} {\bibfnamefont {V.}~\bibnamefont {Smelyanskiy}},\ }\href {\doibase 10.1103/PhysRevApplied.20.054008} {\bibfield  {journal} {\bibinfo  {journal} {Phys. Rev. Appl.}\ }\textbf {\bibinfo {volume} {20}},\ \bibinfo {pages} {054008} (\bibinfo {year} {2023})}\BibitemShut {NoStop}%
\bibitem [{\citenamefont {Dumas}\ \emph {et~al.}(2024)\citenamefont {Dumas}, \citenamefont {Groleau-Par\'e}, \citenamefont {McDonald}, \citenamefont {Mu\~noz Arias}, \citenamefont {Lled\'o}, \citenamefont {D'Anjou},\ and\ \citenamefont {Blais}}]{unified_ionization}%
  \BibitemOpen
  \bibfield  {author} {\bibinfo {author} {\bibfnamefont {M.~F.}\ \bibnamefont {Dumas}}, \bibinfo {author} {\bibfnamefont {B.}~\bibnamefont {Groleau-Par\'e}}, \bibinfo {author} {\bibfnamefont {A.}~\bibnamefont {McDonald}}, \bibinfo {author} {\bibfnamefont {M.~H.}\ \bibnamefont {Mu\~noz Arias}}, \bibinfo {author} {\bibfnamefont {C.}~\bibnamefont {Lled\'o}}, \bibinfo {author} {\bibfnamefont {B.}~\bibnamefont {D'Anjou}}, \ and\ \bibinfo {author} {\bibfnamefont {A.}~\bibnamefont {Blais}},\ }\href {\doibase 10.1103/PhysRevX.14.041023} {\bibfield  {journal} {\bibinfo  {journal} {Phys. Rev. X}\ }\textbf {\bibinfo {volume} {14}},\ \bibinfo {pages} {041023} (\bibinfo {year} {2024})}\BibitemShut {NoStop}%
\bibitem [{\citenamefont {Ghosh}\ \emph {et~al.}(2013)\citenamefont {Ghosh}, \citenamefont {Fowler}, \citenamefont {Martinis},\ and\ \citenamefont {Geller}}]{Ghosh_2013}%
  \BibitemOpen
  \bibfield  {author} {\bibinfo {author} {\bibfnamefont {J.}~\bibnamefont {Ghosh}}, \bibinfo {author} {\bibfnamefont {A.~G.}\ \bibnamefont {Fowler}}, \bibinfo {author} {\bibfnamefont {J.~M.}\ \bibnamefont {Martinis}}, \ and\ \bibinfo {author} {\bibfnamefont {M.~R.}\ \bibnamefont {Geller}},\ }\href {\doibase 10.1103/PhysRevA.88.062329} {\bibfield  {journal} {\bibinfo  {journal} {Phys. Rev. A}\ }\textbf {\bibinfo {volume} {88}},\ \bibinfo {pages} {062329} (\bibinfo {year} {2013})}\BibitemShut {NoStop}%
\bibitem [{\citenamefont {Fowler}(2013)}]{Fowler_2013}%
  \BibitemOpen
  \bibfield  {author} {\bibinfo {author} {\bibfnamefont {A.~G.}\ \bibnamefont {Fowler}},\ }\href {\doibase 10.1103/PhysRevA.88.042308} {\bibfield  {journal} {\bibinfo  {journal} {Phys. Rev. A}\ }\textbf {\bibinfo {volume} {88}},\ \bibinfo {pages} {042308} (\bibinfo {year} {2013})}\BibitemShut {NoStop}%
\bibitem [{\citenamefont {Varbanov}\ \emph {et~al.}(2020)\citenamefont {Varbanov}, \citenamefont {Battistel}, \citenamefont {Tarasinski}, \citenamefont {Ostroukh}, \citenamefont {O’Brien}, \citenamefont {DiCarlo},\ and\ \citenamefont {Terhal}}]{varbanov2020leakage}%
  \BibitemOpen
  \bibfield  {author} {\bibinfo {author} {\bibfnamefont {B.~M.}\ \bibnamefont {Varbanov}}, \bibinfo {author} {\bibfnamefont {F.}~\bibnamefont {Battistel}}, \bibinfo {author} {\bibfnamefont {B.~M.}\ \bibnamefont {Tarasinski}}, \bibinfo {author} {\bibfnamefont {V.~P.}\ \bibnamefont {Ostroukh}}, \bibinfo {author} {\bibfnamefont {T.~E.}\ \bibnamefont {O’Brien}}, \bibinfo {author} {\bibfnamefont {L.}~\bibnamefont {DiCarlo}}, \ and\ \bibinfo {author} {\bibfnamefont {B.~M.}\ \bibnamefont {Terhal}},\ }\href@noop {} {\bibfield  {journal} {\bibinfo  {journal} {npj Quantum Information}\ }\textbf {\bibinfo {volume} {6}},\ \bibinfo {pages} {102} (\bibinfo {year} {2020})}\BibitemShut {NoStop}%
\bibitem [{\citenamefont {Miao}\ \emph {et~al.}(2023)\citenamefont {Miao}, \citenamefont {McEwen}, \citenamefont {Atalaya}, \citenamefont {Kafri}, \citenamefont {Pryadko}, \citenamefont {Bengtsson}, \citenamefont {Opremcak}, \citenamefont {Satzinger}, \citenamefont {Chen}, \citenamefont {Klimov} \emph {et~al.}}]{miao_2023}%
  \BibitemOpen
  \bibfield  {author} {\bibinfo {author} {\bibfnamefont {K.~C.}\ \bibnamefont {Miao}}, \bibinfo {author} {\bibfnamefont {M.}~\bibnamefont {McEwen}}, \bibinfo {author} {\bibfnamefont {J.}~\bibnamefont {Atalaya}}, \bibinfo {author} {\bibfnamefont {D.}~\bibnamefont {Kafri}}, \bibinfo {author} {\bibfnamefont {L.~P.}\ \bibnamefont {Pryadko}}, \bibinfo {author} {\bibfnamefont {A.}~\bibnamefont {Bengtsson}}, \bibinfo {author} {\bibfnamefont {A.}~\bibnamefont {Opremcak}}, \bibinfo {author} {\bibfnamefont {K.~J.}\ \bibnamefont {Satzinger}}, \bibinfo {author} {\bibfnamefont {Z.}~\bibnamefont {Chen}}, \bibinfo {author} {\bibfnamefont {P.~V.}\ \bibnamefont {Klimov}},  \emph {et~al.},\ }\href@noop {} {\bibfield  {journal} {\bibinfo  {journal} {Nature Physics}\ }\textbf {\bibinfo {volume} {19}},\ \bibinfo {pages} {1780} (\bibinfo {year} {2023})}\BibitemShut {NoStop}%
\bibitem [{sup()}]{supp}%
  \BibitemOpen
  \href@noop {} {}\bibinfo {note} {See Supplemental Material for additional information and data of the experiments.}\BibitemShut {Stop}%
\bibitem [{\citenamefont {Jeffrey}\ \emph {et~al.}(2014)\citenamefont {Jeffrey}, \citenamefont {Sank}, \citenamefont {Mutus}, \citenamefont {White}, \citenamefont {Kelly}, \citenamefont {Barends}, \citenamefont {Chen}, \citenamefont {Chen}, \citenamefont {Chiaro}, \citenamefont {Dunsworth} \emph {et~al.}}]{jeffrey_2014}%
  \BibitemOpen
  \bibfield  {author} {\bibinfo {author} {\bibfnamefont {E.}~\bibnamefont {Jeffrey}}, \bibinfo {author} {\bibfnamefont {D.}~\bibnamefont {Sank}}, \bibinfo {author} {\bibfnamefont {J.~Y.}\ \bibnamefont {Mutus}}, \bibinfo {author} {\bibfnamefont {T.~C.}\ \bibnamefont {White}}, \bibinfo {author} {\bibfnamefont {J.}~\bibnamefont {Kelly}}, \bibinfo {author} {\bibfnamefont {R.}~\bibnamefont {Barends}}, \bibinfo {author} {\bibfnamefont {Y.}~\bibnamefont {Chen}}, \bibinfo {author} {\bibfnamefont {Z.}~\bibnamefont {Chen}}, \bibinfo {author} {\bibfnamefont {B.}~\bibnamefont {Chiaro}}, \bibinfo {author} {\bibfnamefont {A.}~\bibnamefont {Dunsworth}},  \emph {et~al.},\ }\href {\doibase 10.1103/PhysRevLett.112.190504} {\bibfield  {journal} {\bibinfo  {journal} {Phys. Rev. Lett.}\ }\textbf {\bibinfo {volume} {112}},\ \bibinfo {pages} {190504} (\bibinfo {year} {2014})}\BibitemShut {NoStop}%
\bibitem [{\citenamefont {Bultink}\ \emph {et~al.}(2016)\citenamefont {Bultink}, \citenamefont {Rol}, \citenamefont {O'Brien}, \citenamefont {Fu}, \citenamefont {Dikken}, \citenamefont {Dickel}, \citenamefont {Vermeulen}, \citenamefont {de~Sterke}, \citenamefont {Bruno}, \citenamefont {Schouten},\ and\ \citenamefont {DiCarlo}}]{bultink_2016}%
  \BibitemOpen
  \bibfield  {author} {\bibinfo {author} {\bibfnamefont {C.~C.}\ \bibnamefont {Bultink}}, \bibinfo {author} {\bibfnamefont {M.~A.}\ \bibnamefont {Rol}}, \bibinfo {author} {\bibfnamefont {T.~E.}\ \bibnamefont {O'Brien}}, \bibinfo {author} {\bibfnamefont {X.}~\bibnamefont {Fu}}, \bibinfo {author} {\bibfnamefont {B.~C.~S.}\ \bibnamefont {Dikken}}, \bibinfo {author} {\bibfnamefont {C.}~\bibnamefont {Dickel}}, \bibinfo {author} {\bibfnamefont {R.~F.~L.}\ \bibnamefont {Vermeulen}}, \bibinfo {author} {\bibfnamefont {J.~C.}\ \bibnamefont {de~Sterke}}, \bibinfo {author} {\bibfnamefont {A.}~\bibnamefont {Bruno}}, \bibinfo {author} {\bibfnamefont {R.~N.}\ \bibnamefont {Schouten}}, \ and\ \bibinfo {author} {\bibfnamefont {L.}~\bibnamefont {DiCarlo}},\ }\href {\doibase 10.1103/PhysRevApplied.6.034008} {\bibfield  {journal} {\bibinfo  {journal} {Phys. Rev. Appl.}\ }\textbf {\bibinfo {volume} {6}},\ \bibinfo {pages} {034008} (\bibinfo {year} {2016})}\BibitemShut {NoStop}%
\bibitem [{\citenamefont {Walter}\ \emph {et~al.}(2017)\citenamefont {Walter}, \citenamefont {Kurpiers}, \citenamefont {Gasparinetti}, \citenamefont {Magnard}, \citenamefont {Poto\ifmmode~\check{c}\else \v{c}\fi{}nik}, \citenamefont {Salath\'e}, \citenamefont {Pechal}, \citenamefont {Mondal}, \citenamefont {Oppliger}, \citenamefont {Eichler},\ and\ \citenamefont {Wallraff}}]{walter_2017}%
  \BibitemOpen
  \bibfield  {author} {\bibinfo {author} {\bibfnamefont {T.}~\bibnamefont {Walter}}, \bibinfo {author} {\bibfnamefont {P.}~\bibnamefont {Kurpiers}}, \bibinfo {author} {\bibfnamefont {S.}~\bibnamefont {Gasparinetti}}, \bibinfo {author} {\bibfnamefont {P.}~\bibnamefont {Magnard}}, \bibinfo {author} {\bibfnamefont {A.}~\bibnamefont {Poto\ifmmode~\check{c}\else \v{c}\fi{}nik}}, \bibinfo {author} {\bibfnamefont {Y.}~\bibnamefont {Salath\'e}}, \bibinfo {author} {\bibfnamefont {M.}~\bibnamefont {Pechal}}, \bibinfo {author} {\bibfnamefont {M.}~\bibnamefont {Mondal}}, \bibinfo {author} {\bibfnamefont {M.}~\bibnamefont {Oppliger}}, \bibinfo {author} {\bibfnamefont {C.}~\bibnamefont {Eichler}}, \ and\ \bibinfo {author} {\bibfnamefont {A.}~\bibnamefont {Wallraff}},\ }\href {\doibase 10.1103/PhysRevApplied.7.054020} {\bibfield  {journal} {\bibinfo  {journal} {Phys. Rev. Appl.}\ }\textbf {\bibinfo {volume} {7}},\ \bibinfo {pages} {054020} (\bibinfo {year} {2017})}\BibitemShut {NoStop}%
\bibitem [{\citenamefont {Touzard}\ \emph {et~al.}(2019)\citenamefont {Touzard}, \citenamefont {Kou}, \citenamefont {Frattini}, \citenamefont {Sivak}, \citenamefont {Puri}, \citenamefont {Grimm}, \citenamefont {Frunzio}, \citenamefont {Shankar},\ and\ \citenamefont {Devoret}}]{touzard_cd}%
  \BibitemOpen
  \bibfield  {author} {\bibinfo {author} {\bibfnamefont {S.}~\bibnamefont {Touzard}}, \bibinfo {author} {\bibfnamefont {A.}~\bibnamefont {Kou}}, \bibinfo {author} {\bibfnamefont {N.~E.}\ \bibnamefont {Frattini}}, \bibinfo {author} {\bibfnamefont {V.~V.}\ \bibnamefont {Sivak}}, \bibinfo {author} {\bibfnamefont {S.}~\bibnamefont {Puri}}, \bibinfo {author} {\bibfnamefont {A.}~\bibnamefont {Grimm}}, \bibinfo {author} {\bibfnamefont {L.}~\bibnamefont {Frunzio}}, \bibinfo {author} {\bibfnamefont {S.}~\bibnamefont {Shankar}}, \ and\ \bibinfo {author} {\bibfnamefont {M.~H.}\ \bibnamefont {Devoret}},\ }\href {\doibase 10.1103/PhysRevLett.122.080502} {\bibfield  {journal} {\bibinfo  {journal} {Phys. Rev. Lett.}\ }\textbf {\bibinfo {volume} {122}},\ \bibinfo {pages} {080502} (\bibinfo {year} {2019})}\BibitemShut {NoStop}%
\bibitem [{\citenamefont {Dassonneville}\ \emph {et~al.}(2020)\citenamefont {Dassonneville}, \citenamefont {Ramos}, \citenamefont {Milchakov}, \citenamefont {Planat}, \citenamefont {Dumur}, \citenamefont {Foroughi}, \citenamefont {Puertas}, \citenamefont {Leger}, \citenamefont {Bharadwaj}, \citenamefont {Delaforce}, \citenamefont {Naud}, \citenamefont {Hasch-Guichard}, \citenamefont {Garc\'{\i}a-Ripoll}, \citenamefont {Roch},\ and\ \citenamefont {Buisson}}]{dassonneville_2020}%
  \BibitemOpen
  \bibfield  {author} {\bibinfo {author} {\bibfnamefont {R.}~\bibnamefont {Dassonneville}}, \bibinfo {author} {\bibfnamefont {T.}~\bibnamefont {Ramos}}, \bibinfo {author} {\bibfnamefont {V.}~\bibnamefont {Milchakov}}, \bibinfo {author} {\bibfnamefont {L.}~\bibnamefont {Planat}}, \bibinfo {author} {\bibfnamefont {E.}~\bibnamefont {Dumur}}, \bibinfo {author} {\bibfnamefont {F.}~\bibnamefont {Foroughi}}, \bibinfo {author} {\bibfnamefont {J.}~\bibnamefont {Puertas}}, \bibinfo {author} {\bibfnamefont {S.}~\bibnamefont {Leger}}, \bibinfo {author} {\bibfnamefont {K.}~\bibnamefont {Bharadwaj}}, \bibinfo {author} {\bibfnamefont {J.}~\bibnamefont {Delaforce}}, \bibinfo {author} {\bibfnamefont {C.}~\bibnamefont {Naud}}, \bibinfo {author} {\bibfnamefont {W.}~\bibnamefont {Hasch-Guichard}}, \bibinfo {author} {\bibfnamefont {J.~J.}\ \bibnamefont {Garc\'{\i}a-Ripoll}}, \bibinfo {author} {\bibfnamefont {N.}~\bibnamefont {Roch}}, \ and\ \bibinfo {author} {\bibfnamefont {O.}~\bibnamefont {Buisson}},\ }\href {\doibase
  10.1103/PhysRevX.10.011045} {\bibfield  {journal} {\bibinfo  {journal} {Phys. Rev. X}\ }\textbf {\bibinfo {volume} {10}},\ \bibinfo {pages} {011045} (\bibinfo {year} {2020})}\BibitemShut {NoStop}%
\bibitem [{\citenamefont {Sunada}\ \emph {et~al.}(2022)\citenamefont {Sunada}, \citenamefont {Kono}, \citenamefont {Ilves}, \citenamefont {Tamate}, \citenamefont {Sugiyama}, \citenamefont {Tabuchi},\ and\ \citenamefont {Nakamura}}]{sunada_2022}%
  \BibitemOpen
  \bibfield  {author} {\bibinfo {author} {\bibfnamefont {Y.}~\bibnamefont {Sunada}}, \bibinfo {author} {\bibfnamefont {S.}~\bibnamefont {Kono}}, \bibinfo {author} {\bibfnamefont {J.}~\bibnamefont {Ilves}}, \bibinfo {author} {\bibfnamefont {S.}~\bibnamefont {Tamate}}, \bibinfo {author} {\bibfnamefont {T.}~\bibnamefont {Sugiyama}}, \bibinfo {author} {\bibfnamefont {Y.}~\bibnamefont {Tabuchi}}, \ and\ \bibinfo {author} {\bibfnamefont {Y.}~\bibnamefont {Nakamura}},\ }\href {\doibase 10.1103/PhysRevApplied.17.044016} {\bibfield  {journal} {\bibinfo  {journal} {Phys. Rev. Appl.}\ }\textbf {\bibinfo {volume} {17}},\ \bibinfo {pages} {044016} (\bibinfo {year} {2022})}\BibitemShut {NoStop}%
\bibitem [{\citenamefont {Braginsky}\ \emph {et~al.}(1980)\citenamefont {Braginsky}, \citenamefont {Vorontsov},\ and\ \citenamefont {Thorne}}]{Braginsky_qnd}%
  \BibitemOpen
  \bibfield  {author} {\bibinfo {author} {\bibfnamefont {V.~B.}\ \bibnamefont {Braginsky}}, \bibinfo {author} {\bibfnamefont {Y.~I.}\ \bibnamefont {Vorontsov}}, \ and\ \bibinfo {author} {\bibfnamefont {K.~S.}\ \bibnamefont {Thorne}},\ }\href {\doibase 10.1126/science.209.4456.547} {\bibfield  {journal} {\bibinfo  {journal} {Science}\ }\textbf {\bibinfo {volume} {209}},\ \bibinfo {pages} {547} (\bibinfo {year} {1980})}\BibitemShut {NoStop}%
\bibitem [{\citenamefont {Caves}\ \emph {et~al.}(1980)\citenamefont {Caves}, \citenamefont {Thorne}, \citenamefont {Drever}, \citenamefont {Sandberg},\ and\ \citenamefont {Zimmermann}}]{caves_qnd}%
  \BibitemOpen
  \bibfield  {author} {\bibinfo {author} {\bibfnamefont {C.~M.}\ \bibnamefont {Caves}}, \bibinfo {author} {\bibfnamefont {K.~S.}\ \bibnamefont {Thorne}}, \bibinfo {author} {\bibfnamefont {R.~W.~P.}\ \bibnamefont {Drever}}, \bibinfo {author} {\bibfnamefont {V.~D.}\ \bibnamefont {Sandberg}}, \ and\ \bibinfo {author} {\bibfnamefont {M.}~\bibnamefont {Zimmermann}},\ }\href {\doibase 10.1103/RevModPhys.52.341} {\bibfield  {journal} {\bibinfo  {journal} {Rev. Mod. Phys.}\ }\textbf {\bibinfo {volume} {52}},\ \bibinfo {pages} {341} (\bibinfo {year} {1980})}\BibitemShut {NoStop}%
\bibitem [{\citenamefont {Braginsky}\ and\ \citenamefont {Khalili}(1996)}]{qnd_review}%
  \BibitemOpen
  \bibfield  {author} {\bibinfo {author} {\bibfnamefont {V.~B.}\ \bibnamefont {Braginsky}}\ and\ \bibinfo {author} {\bibfnamefont {F.~Y.}\ \bibnamefont {Khalili}},\ }\href {\doibase 10.1103/RevModPhys.68.1} {\bibfield  {journal} {\bibinfo  {journal} {Rev. Mod. Phys.}\ }\textbf {\bibinfo {volume} {68}},\ \bibinfo {pages} {1} (\bibinfo {year} {1996})}\BibitemShut {NoStop}%
\bibitem [{\citenamefont {Spring}\ \emph {et~al.}(2024)\citenamefont {Spring}, \citenamefont {Milanovic}, \citenamefont {Sunada}, \citenamefont {Wang}, \citenamefont {van Loo}, \citenamefont {Tamate},\ and\ \citenamefont {Nakamura}}]{spring_multiplexed_2024}%
  \BibitemOpen
  \bibfield  {author} {\bibinfo {author} {\bibfnamefont {P.~A.}\ \bibnamefont {Spring}}, \bibinfo {author} {\bibfnamefont {L.}~\bibnamefont {Milanovic}}, \bibinfo {author} {\bibfnamefont {Y.}~\bibnamefont {Sunada}}, \bibinfo {author} {\bibfnamefont {S.}~\bibnamefont {Wang}}, \bibinfo {author} {\bibfnamefont {A.~F.}\ \bibnamefont {van Loo}}, \bibinfo {author} {\bibfnamefont {S.}~\bibnamefont {Tamate}}, \ and\ \bibinfo {author} {\bibfnamefont {Y.}~\bibnamefont {Nakamura}},\ }\href@noop {} {\bibfield  {journal} {\bibinfo  {journal} {arXiv preprint arXiv:2409.04967}\ } (\bibinfo {year} {2024})}\BibitemShut {NoStop}%
\bibitem [{\citenamefont {Chen}\ \emph {et~al.}(2023)\citenamefont {Chen}, \citenamefont {Li}, \citenamefont {Lu}, \citenamefont {Warren}, \citenamefont {Kri{\v{z}}an}, \citenamefont {Kosen}, \citenamefont {Rommel}, \citenamefont {Ahmed}, \citenamefont {Osman}, \citenamefont {Bizn{\'a}rov{\'a}} \emph {et~al.}}]{chen_transmon}%
  \BibitemOpen
  \bibfield  {author} {\bibinfo {author} {\bibfnamefont {L.}~\bibnamefont {Chen}}, \bibinfo {author} {\bibfnamefont {H.-X.}\ \bibnamefont {Li}}, \bibinfo {author} {\bibfnamefont {Y.}~\bibnamefont {Lu}}, \bibinfo {author} {\bibfnamefont {C.~W.}\ \bibnamefont {Warren}}, \bibinfo {author} {\bibfnamefont {C.~J.}\ \bibnamefont {Kri{\v{z}}an}}, \bibinfo {author} {\bibfnamefont {S.}~\bibnamefont {Kosen}}, \bibinfo {author} {\bibfnamefont {M.}~\bibnamefont {Rommel}}, \bibinfo {author} {\bibfnamefont {S.}~\bibnamefont {Ahmed}}, \bibinfo {author} {\bibfnamefont {A.}~\bibnamefont {Osman}}, \bibinfo {author} {\bibfnamefont {J.}~\bibnamefont {Bizn{\'a}rov{\'a}}},  \emph {et~al.},\ }\href@noop {} {\bibfield  {journal} {\bibinfo  {journal} {npj Quantum Information}\ }\textbf {\bibinfo {volume} {9}},\ \bibinfo {pages} {26} (\bibinfo {year} {2023})}\BibitemShut {NoStop}%
\bibitem [{\citenamefont {Heinsoo}\ \emph {et~al.}(2018)\citenamefont {Heinsoo}, \citenamefont {Andersen}, \citenamefont {Remm}, \citenamefont {Krinner}, \citenamefont {Walter}, \citenamefont {Salath\'e}, \citenamefont {Gasparinetti}, \citenamefont {Besse}, \citenamefont {Poto\ifmmode~\check{c}\else \v{c}\fi{}nik}, \citenamefont {Wallraff},\ and\ \citenamefont {Eichler}}]{multiplexed_eth}%
  \BibitemOpen
  \bibfield  {author} {\bibinfo {author} {\bibfnamefont {J.}~\bibnamefont {Heinsoo}}, \bibinfo {author} {\bibfnamefont {C.~K.}\ \bibnamefont {Andersen}}, \bibinfo {author} {\bibfnamefont {A.}~\bibnamefont {Remm}}, \bibinfo {author} {\bibfnamefont {S.}~\bibnamefont {Krinner}}, \bibinfo {author} {\bibfnamefont {T.}~\bibnamefont {Walter}}, \bibinfo {author} {\bibfnamefont {Y.}~\bibnamefont {Salath\'e}}, \bibinfo {author} {\bibfnamefont {S.}~\bibnamefont {Gasparinetti}}, \bibinfo {author} {\bibfnamefont {J.-C.}\ \bibnamefont {Besse}}, \bibinfo {author} {\bibfnamefont {A.}~\bibnamefont {Poto\ifmmode~\check{c}\else \v{c}\fi{}nik}}, \bibinfo {author} {\bibfnamefont {A.}~\bibnamefont {Wallraff}}, \ and\ \bibinfo {author} {\bibfnamefont {C.}~\bibnamefont {Eichler}},\ }\href {\doibase 10.1103/PhysRevApplied.10.034040} {\bibfield  {journal} {\bibinfo  {journal} {Phys. Rev. Appl.}\ }\textbf {\bibinfo {volume} {10}},\ \bibinfo {pages} {034040} (\bibinfo {year} {2018})}\BibitemShut {NoStop}%
\bibitem [{\citenamefont {Gambetta}\ \emph {et~al.}(2011)\citenamefont {Gambetta}, \citenamefont {Houck},\ and\ \citenamefont {Blais}}]{gambetta_2011}%
  \BibitemOpen
  \bibfield  {author} {\bibinfo {author} {\bibfnamefont {J.~M.}\ \bibnamefont {Gambetta}}, \bibinfo {author} {\bibfnamefont {A.~A.}\ \bibnamefont {Houck}}, \ and\ \bibinfo {author} {\bibfnamefont {A.}~\bibnamefont {Blais}},\ }\href {\doibase 10.1103/PhysRevLett.106.030502} {\bibfield  {journal} {\bibinfo  {journal} {Phys. Rev. Lett.}\ }\textbf {\bibinfo {volume} {106}},\ \bibinfo {pages} {030502} (\bibinfo {year} {2011})}\BibitemShut {NoStop}%
\bibitem [{\citenamefont {Diniz}\ \emph {et~al.}(2013)\citenamefont {Diniz}, \citenamefont {Dumur}, \citenamefont {Buisson},\ and\ \citenamefont {Auff\`eves}}]{diniz_2013}%
  \BibitemOpen
  \bibfield  {author} {\bibinfo {author} {\bibfnamefont {I.}~\bibnamefont {Diniz}}, \bibinfo {author} {\bibfnamefont {E.}~\bibnamefont {Dumur}}, \bibinfo {author} {\bibfnamefont {O.}~\bibnamefont {Buisson}}, \ and\ \bibinfo {author} {\bibfnamefont {A.}~\bibnamefont {Auff\`eves}},\ }\href {\doibase 10.1103/PhysRevA.87.033837} {\bibfield  {journal} {\bibinfo  {journal} {Phys. Rev. A}\ }\textbf {\bibinfo {volume} {87}},\ \bibinfo {pages} {033837} (\bibinfo {year} {2013})}\BibitemShut {NoStop}%
\bibitem [{\citenamefont {Roy}\ \emph {et~al.}(2017)\citenamefont {Roy}, \citenamefont {Kundu}, \citenamefont {Chand}, \citenamefont {Hazra}, \citenamefont {Nehra}, \citenamefont {Cosmic}, \citenamefont {Ranadive}, \citenamefont {Patankar}, \citenamefont {Damle},\ and\ \citenamefont {Vijay}}]{roy_2017}%
  \BibitemOpen
  \bibfield  {author} {\bibinfo {author} {\bibfnamefont {T.}~\bibnamefont {Roy}}, \bibinfo {author} {\bibfnamefont {S.}~\bibnamefont {Kundu}}, \bibinfo {author} {\bibfnamefont {M.}~\bibnamefont {Chand}}, \bibinfo {author} {\bibfnamefont {S.}~\bibnamefont {Hazra}}, \bibinfo {author} {\bibfnamefont {N.}~\bibnamefont {Nehra}}, \bibinfo {author} {\bibfnamefont {R.}~\bibnamefont {Cosmic}}, \bibinfo {author} {\bibfnamefont {A.}~\bibnamefont {Ranadive}}, \bibinfo {author} {\bibfnamefont {M.~P.}\ \bibnamefont {Patankar}}, \bibinfo {author} {\bibfnamefont {K.}~\bibnamefont {Damle}}, \ and\ \bibinfo {author} {\bibfnamefont {R.}~\bibnamefont {Vijay}},\ }\href {\doibase 10.1103/PhysRevApplied.7.054025} {\bibfield  {journal} {\bibinfo  {journal} {Phys. Rev. Appl.}\ }\textbf {\bibinfo {volume} {7}},\ \bibinfo {pages} {054025} (\bibinfo {year} {2017})}\BibitemShut {NoStop}%
\bibitem [{\citenamefont {Pfeiffer}\ \emph {et~al.}(2024)\citenamefont {Pfeiffer}, \citenamefont {Werninghaus}, \citenamefont {Schweizer}, \citenamefont {Bruckmoser}, \citenamefont {Koch}, \citenamefont {Glaser}, \citenamefont {Huber}, \citenamefont {Bunch}, \citenamefont {Haslbeck}, \citenamefont {Knudsen}, \citenamefont {Krylov}, \citenamefont {Liegener}, \citenamefont {Marx}, \citenamefont {Richard}, \citenamefont {Romeiro}, \citenamefont {Roy}, \citenamefont {Schirk}, \citenamefont {Schneider}, \citenamefont {Singh}, \citenamefont {S\"odergren}, \citenamefont {Tsitsilin}, \citenamefont {Wallner}, \citenamefont {Riofr\'{\i}o},\ and\ \citenamefont {Filipp}}]{pfeiffer_2024}%
  \BibitemOpen
  \bibfield  {author} {\bibinfo {author} {\bibfnamefont {F.}~\bibnamefont {Pfeiffer}}, \bibinfo {author} {\bibfnamefont {M.}~\bibnamefont {Werninghaus}}, \bibinfo {author} {\bibfnamefont {C.}~\bibnamefont {Schweizer}}, \bibinfo {author} {\bibfnamefont {N.}~\bibnamefont {Bruckmoser}}, \bibinfo {author} {\bibfnamefont {L.}~\bibnamefont {Koch}}, \bibinfo {author} {\bibfnamefont {N.~J.}\ \bibnamefont {Glaser}}, \bibinfo {author} {\bibfnamefont {G.~B.~P.}\ \bibnamefont {Huber}}, \bibinfo {author} {\bibfnamefont {D.}~\bibnamefont {Bunch}}, \bibinfo {author} {\bibfnamefont {F.~X.}\ \bibnamefont {Haslbeck}}, \bibinfo {author} {\bibfnamefont {M.}~\bibnamefont {Knudsen}}, \bibinfo {author} {\bibfnamefont {G.}~\bibnamefont {Krylov}}, \bibinfo {author} {\bibfnamefont {K.}~\bibnamefont {Liegener}}, \bibinfo {author} {\bibfnamefont {A.}~\bibnamefont {Marx}}, \bibinfo {author} {\bibfnamefont {L.}~\bibnamefont {Richard}}, \bibinfo {author} {\bibfnamefont {J.~H.}\ \bibnamefont {Romeiro}}, \bibinfo {author} {\bibfnamefont
  {F.~A.}\ \bibnamefont {Roy}}, \bibinfo {author} {\bibfnamefont {J.}~\bibnamefont {Schirk}}, \bibinfo {author} {\bibfnamefont {C.}~\bibnamefont {Schneider}}, \bibinfo {author} {\bibfnamefont {M.}~\bibnamefont {Singh}}, \bibinfo {author} {\bibfnamefont {L.}~\bibnamefont {S\"odergren}}, \bibinfo {author} {\bibfnamefont {I.}~\bibnamefont {Tsitsilin}}, \bibinfo {author} {\bibfnamefont {F.}~\bibnamefont {Wallner}}, \bibinfo {author} {\bibfnamefont {C.~A.}\ \bibnamefont {Riofr\'{\i}o}}, \ and\ \bibinfo {author} {\bibfnamefont {S.}~\bibnamefont {Filipp}},\ }\href {\doibase 10.1103/PhysRevX.14.041007} {\bibfield  {journal} {\bibinfo  {journal} {Phys. Rev. X}\ }\textbf {\bibinfo {volume} {14}},\ \bibinfo {pages} {041007} (\bibinfo {year} {2024})}\BibitemShut {NoStop}%
\bibitem [{\citenamefont {Wood}\ and\ \citenamefont {Gambetta}(2018)}]{wood_lrb_2018}%
  \BibitemOpen
  \bibfield  {author} {\bibinfo {author} {\bibfnamefont {C.~J.}\ \bibnamefont {Wood}}\ and\ \bibinfo {author} {\bibfnamefont {J.~M.}\ \bibnamefont {Gambetta}},\ }\href {\doibase 10.1103/PhysRevA.97.032306} {\bibfield  {journal} {\bibinfo  {journal} {Phys. Rev. A}\ }\textbf {\bibinfo {volume} {97}},\ \bibinfo {pages} {032306} (\bibinfo {year} {2018})}\BibitemShut {NoStop}%
\bibitem [{\citenamefont {Wallraff}\ \emph {et~al.}(2004)\citenamefont {Wallraff}, \citenamefont {Schuster}, \citenamefont {Blais}, \citenamefont {Frunzio}, \citenamefont {Huang}, \citenamefont {Majer}, \citenamefont {Kumar}, \citenamefont {Girvin},\ and\ \citenamefont {Schoelkopf}}]{wallraff_disp_2004}%
  \BibitemOpen
  \bibfield  {author} {\bibinfo {author} {\bibfnamefont {A.}~\bibnamefont {Wallraff}}, \bibinfo {author} {\bibfnamefont {D.~I.}\ \bibnamefont {Schuster}}, \bibinfo {author} {\bibfnamefont {A.}~\bibnamefont {Blais}}, \bibinfo {author} {\bibfnamefont {L.}~\bibnamefont {Frunzio}}, \bibinfo {author} {\bibfnamefont {R.-S.}\ \bibnamefont {Huang}}, \bibinfo {author} {\bibfnamefont {J.}~\bibnamefont {Majer}}, \bibinfo {author} {\bibfnamefont {S.}~\bibnamefont {Kumar}}, \bibinfo {author} {\bibfnamefont {S.~M.}\ \bibnamefont {Girvin}}, \ and\ \bibinfo {author} {\bibfnamefont {R.~J.}\ \bibnamefont {Schoelkopf}},\ }\href {\doibase 10.1038/nature02851} {\bibfield  {journal} {\bibinfo  {journal} {Nature}\ }\textbf {\bibinfo {volume} {431}},\ \bibinfo {pages} {162} (\bibinfo {year} {2004})},\ \bibinfo {note} {publisher: Nature Publishing Group}\BibitemShut {NoStop}%
\bibitem [{\citenamefont {McClure}\ \emph {et~al.}(2016)\citenamefont {McClure}, \citenamefont {Paik}, \citenamefont {Bishop}, \citenamefont {Steffen}, \citenamefont {Chow},\ and\ \citenamefont {Gambetta}}]{mcclure_2016}%
  \BibitemOpen
  \bibfield  {author} {\bibinfo {author} {\bibfnamefont {D.~T.}\ \bibnamefont {McClure}}, \bibinfo {author} {\bibfnamefont {H.}~\bibnamefont {Paik}}, \bibinfo {author} {\bibfnamefont {L.~S.}\ \bibnamefont {Bishop}}, \bibinfo {author} {\bibfnamefont {M.}~\bibnamefont {Steffen}}, \bibinfo {author} {\bibfnamefont {J.~M.}\ \bibnamefont {Chow}}, \ and\ \bibinfo {author} {\bibfnamefont {J.~M.}\ \bibnamefont {Gambetta}},\ }\href {\doibase 10.1103/PhysRevApplied.5.011001} {\bibfield  {journal} {\bibinfo  {journal} {Phys. Rev. Appl.}\ }\textbf {\bibinfo {volume} {5}},\ \bibinfo {pages} {011001} (\bibinfo {year} {2016})}\BibitemShut {NoStop}%
\bibitem [{\citenamefont {Frattini}\ \emph {et~al.}(2018)\citenamefont {Frattini}, \citenamefont {Sivak}, \citenamefont {Lingenfelter}, \citenamefont {Shankar},\ and\ \citenamefont {Devoret}}]{spa_opt}%
  \BibitemOpen
  \bibfield  {author} {\bibinfo {author} {\bibfnamefont {N.~E.}\ \bibnamefont {Frattini}}, \bibinfo {author} {\bibfnamefont {V.~V.}\ \bibnamefont {Sivak}}, \bibinfo {author} {\bibfnamefont {A.}~\bibnamefont {Lingenfelter}}, \bibinfo {author} {\bibfnamefont {S.}~\bibnamefont {Shankar}}, \ and\ \bibinfo {author} {\bibfnamefont {M.~H.}\ \bibnamefont {Devoret}},\ }\href {\doibase 10.1103/PhysRevApplied.10.054020} {\bibfield  {journal} {\bibinfo  {journal} {Phys. Rev. Appl.}\ }\textbf {\bibinfo {volume} {10}},\ \bibinfo {pages} {054020} (\bibinfo {year} {2018})}\BibitemShut {NoStop}%
\bibitem [{\citenamefont {Swiadek}\ \emph {et~al.}(2023)\citenamefont {Swiadek}, \citenamefont {Shillito}, \citenamefont {Magnard}, \citenamefont {Remm}, \citenamefont {Hellings}, \citenamefont {Lacroix}, \citenamefont {Ficheux}, \citenamefont {Zanuz}, \citenamefont {Norris}, \citenamefont {Blais}, \citenamefont {Krinner},\ and\ \citenamefont {Wallraff}}]{swiadek_2023}%
  \BibitemOpen
  \bibfield  {author} {\bibinfo {author} {\bibfnamefont {F.}~\bibnamefont {Swiadek}}, \bibinfo {author} {\bibfnamefont {R.}~\bibnamefont {Shillito}}, \bibinfo {author} {\bibfnamefont {P.}~\bibnamefont {Magnard}}, \bibinfo {author} {\bibfnamefont {A.}~\bibnamefont {Remm}}, \bibinfo {author} {\bibfnamefont {C.}~\bibnamefont {Hellings}}, \bibinfo {author} {\bibfnamefont {N.}~\bibnamefont {Lacroix}}, \bibinfo {author} {\bibfnamefont {Q.}~\bibnamefont {Ficheux}}, \bibinfo {author} {\bibfnamefont {D.~C.}\ \bibnamefont {Zanuz}}, \bibinfo {author} {\bibfnamefont {G.~J.}\ \bibnamefont {Norris}}, \bibinfo {author} {\bibfnamefont {A.}~\bibnamefont {Blais}}, \bibinfo {author} {\bibfnamefont {S.}~\bibnamefont {Krinner}}, \ and\ \bibinfo {author} {\bibfnamefont {A.}~\bibnamefont {Wallraff}},\ }\href@noop {} {\enquote {\bibinfo {title} {Enhancing dispersive readout of superconducting qubits through dynamic control of the dispersive shift: Experiment and theory},}\ } (\bibinfo {year} {2023}),\ \Eprint
  {http://arxiv.org/abs/2307.07765} {arXiv:2307.07765 [quant-ph]} \BibitemShut {NoStop}%
\end{thebibliography}
\end{document}


\title{Supplemental Material for ``Benchmarking the readout of a superconducting qubit for repeated measurements''}
\author{S. Hazra}
\email{sumeru.hazra@yale.edu, wei.dai.wd279@yale.edu}
\thanks{These two authors contributed equally}
\author{W. Dai}
\email{sumeru.hazra@yale.edu, wei.dai.wd279@yale.edu}
\thanks{These two authors contributed equally}
\author{T. Connolly}
\author{P. D. Kurilovich}
\author{Z. Wang}
\author{L. Frunzio}
\author{M. H. Devoret}
\email{michel.devoret@yale.edu}
\affiliation{Department of Applied Physics, Yale University, New Haven, Connecticut 06520, USA\\
and Yale Quantum Institute, Yale University, New Haven, Connecticut 06520, USA}

\maketitle
\tableofcontents
\section{Intrinsically Purcell-protected qubit} 
\label{appendix:device}

\begin{figure}[tbh]
\includegraphics{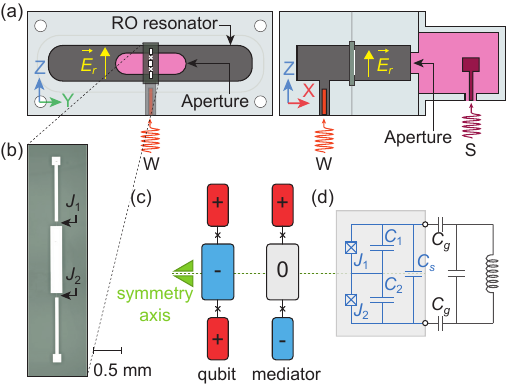}
\caption 
{Experimental device. (a) Schematic of the physical realization of dimon in the 3D architecture viewed as the YZ and the XZ projections.
(b) False-colored optical image of the device. $J_1$ and $J_2$ indicates the two symmetric Josephson junctions (not visible on this scale). 
(c) The two normal modes of a symmetric dimon, termed as the `qubit' and the `mediator' mode. (d) The lumped element circuit model of the dimon coupled to a linear resonator.
}
\label{fig:dimon_device} 
\end{figure}

To eliminate the radiative decay of the qubit we design a symmetric two-mode Josephson device--termed ``dimon''--that benefits from intrinsic Purcell protection~\cite{diniz_2013, roy_2017, dassonneville_2020, pfeiffer_2024}. This circuit was originally proposed as \emph{Tunable Coupling Qubit}~\cite{gambetta_2011} with two SQUIDs in contrast to two Josephson junctions in our implementation.
As illustrated in Fig.~\ref{fig:dimon_device}(a), the device (white), is symmetrically placed at the center of a 3D rectangular cavity (dark gray), functioning as the readout resonator.
The resonator is aperture-coupled to a waveguide filter mode (pink) that strongly emits into a transmission line. 
Such a coupling scheme minimizes parasitic coupling of the dimon to the transmission line.
The bilateral symmetry of the dimon ensures that the two normal modes of the circuit are dipolar and quadrupolar in nature\cite{gambetta_2011}, respectively. 
The dipolar mode undergoes hybridization with the readout resonator through conventional charge-charge linear coupling~\cite{qubit_photon_interaction_dephasing}, whereas the quadrupolar mode has no linear coupling with the readout resonator. We designate the quadrupolar mode as the `qubit', and the dipolar mode as the `mediator'.
In Sec.~\ref{ssec:hamiltonian} we show that the two dimon modes interact via a strong purely non-linear coupling and in Sec.~\ref{ssec:mediated_shift} we explain how it leads to a \textit{mediated} dispersive interaction between the qubit and the readout resonator, despite having no linear hybridization between the two. To the leading order, this mediated dispersive shift depends on the qubit-mediator cross-Kerr interaction $\chi_{qm}$, mediator-readout detuning $\Delta_{mr}$, and their linear coupling strength $g_{mr}$.
The absence of linear hybridization of the qubit with the readout resonator makes  it inherently protected against Purcell decay across all frequencies. Thus, the Purcell protection is robust against Stark shifts induced by all the drives, including the readout drive itself. In Sec.~\ref{ssec:purcell_sim} we consider experimental limitations such as a small asymmetry in the junction fabrication and demonstrate the robustness of the Purcell protection against such practical constraints.

\subsection{Dimon Hamiltonian and energy spectrum}
\label{ssec:hamiltonian}
Here we derive the dimon Hamiltonian, which consists of two Josephson junctions and three capacitors, as schematized in the gray box in Fig. ~\ref{fig:dimon_device}(d). 
We pick the node fluxes $\hat{\Phi}_1$ and $\hat{\Phi}_2$ defined from branch voltage across $C_1$ and $C_2$ (with the center node chosen as the ground) as the dynamical variables. The Lagrangian of the circuit can be written as: 
\begin{align}\label{eq:Lagrangian1}
\hat{\mathcal{L}} = \frac{1}{2}\left[C_{1}\dot{\hat{\Phi}}_{1}^2 + C_{2}\dot{\hat{\Phi}}_{2}^2 + C_{s}(\dot{\hat{\Phi}}_{2}-\dot{\hat{\Phi}}_{1})^2 \right] \\ \nonumber 
+E_{J1}\cos\left( \frac{\hat{\Phi}_{1}}{\phi_0}  \right)
+E_{J2}\cos\left( \frac{\hat{\Phi}_{2}}{\phi_0}  \right)
\end{align}
with $\phi_0 = \hbar/(2e)$ being the reduced flux quantum. The Hamiltonian can be obtained by a Legendre transform on the Lagrangian: 
\begin{align}
\hat{\mathcal{H}} = 
\frac{s_{11}}{2} \hat{Q}_1^2 + \frac{s_{22}}{2} \hat{Q}_2^2 + s_{12} \hat{Q}_1 \hat{Q}_2 \\ \nonumber
- E_{J1}\cos \left( \frac{\hat{\Phi}_{1}}{\phi_0}  \right) - E_{J2} \cos \left( \frac{\hat{\Phi}_{1}}{\phi_0}  \right)
\end{align}
where 
\begin{equation*}
\begin{split}
s_{11} = \frac{C_s + C_2}{C_1 C_2 + C_s (C_1 + C_2)} \\ 
s_{22} = \frac{C_s + C_1}{C_1 C_2 + C_s (C_1 + C_2)} \\
s_{12} = \frac{C_s}{C_1 C_2 + C_s (C_1 + C_2)}
\end{split}
\end{equation*}
are elements of the inverse-capacitance (elastance) matrix.

We include the effect of offset charge by considering external gate voltages $V_{g,1}$ ($V_{g,2}$) acting on node 1 (2) via $C_{g,1}$ ($C_{g,2}$), 
that introduces extra kinetic terms to the variational equation of motion: 
\begin{equation}
\delta \int_{t_i}^{t_f} \left( \hat{\mathcal{L}} + \sum_{i = 1,2} C_{g,i} V_{g,i} (\dot{\hat{\Phi}}_{i} - V_{g,i}) \right) \mathrm{d} t = 0
\end{equation}
Thus they enter the Hamiltonian as displacements on the charge variables: $\hat{Q}_i \rightarrow \hat{Q}_i - Q_{g,i}$ ($i = 1,2$), with $Q_{g,1} = C_{g,1} V_{g,1}$ and $Q_{g,2} = C_{g,2} V_{g,2}$ referred to as offset charges on the two nodes. 

For the interest of this work, the circuit is designed to have the symmetry: $C_{1} = C_{2} = C_{J}$ and $E_{J1} = E_{J2} = E_{J}$. We further denote $\hat{\Phi}_{q} = (\hat{\Phi}_{1} + \hat{\Phi}_{2})/2$ for the \emph{quadrupolar} degree of freedom and $\hat{\Phi}_{m} = (\hat{\Phi}_{1} - \hat{\Phi}_{2})/2$ for the \emph{dipolar} degree of freedom as indicated in Fig.~\ref{fig:dimon_device}(c). Then the Hamiltonian can be reduced to: 
\begin{equation}
\begin{split}
    \hat{\mathcal{H}}_D = \frac{e^2 (\hat{n}_{q} - n_{g,q})^2}{ C_J} + \frac{e^2 (\hat{n}_{m} - n_{g,m})^2}{C_J + 2 C_s} \\ - 2 E_{J}\cos \hat{\varphi}_{q} \cos \hat{\varphi}_{m},
    \label{eq:HD}
\end{split}
\end{equation}
where $\hat{\varphi}_{k} = \hat{\Phi}_{k}/\phi_0$ and $\hat{n}_{k} = \hat{Q}_{k}/2e$ ($k = q, m$) are canonical variables, $n_{g,q} = (Q_{g,1} + Q_{g,2})/2e$ and $n_{g,m} = (Q_{g,1} - Q_{g,2})/2e$ are the effective offset charges viewed by the qubit mode and mediator mode respectively. 

The charging energies
\begin{equation*}
E_{c,q} = \frac{e^2}{2(2C_{J})},  \qquad E_{c,m} = \frac{e^2}{2(2C_{J} + 4 C_{s})}
\end{equation*}
are designed to be small compared $E_J$, ensuring both qubit and mediator modes to be in the transmon regime~\cite{transmon}. As a result, the first few energy levels of dimon are expected to be insensitive to offset charges, which we will neglect for the rest of this work. It is helpful to introduce creation and annihilation operators for the two modes ($k = q, m$): 
\begin{align}
\hat{\varphi}_{k} &= \left( \frac{E_{c,k}}{E_J} \right)^{1/4} (\ad_k + \a_k), \\
\hat{n}_{k} &= \frac{\mathrm{i}}{2} \left( \frac{E_J}{E_{c,k}} \right)^{1/4} (\ad_k - \a_k)
\end{align}

We expand the product of two cosines up to 4th order and keep only the non-rotating terms: 
\begin{align}\hat{\mathcal{H}}_{\mathrm{nl}} &= -2 E_J \left( \cos \hat{\varphi}_{q} \cos \hat{\varphi}_{m} + \frac{\hat{\varphi}_{q}^2 + \hat{\varphi}_{m}^2}{2} \right) \\ \label{eq:H_nl}
&\approx - \frac{E_{c,q}}{2} (\ad_q)^2 \a_q^2 - ( E_{c,q} + \sqrt{E_{c,q} E_{c,m}})\ad_q \a_q \\ \nonumber
& - \frac{E_{c,m}}{2} (\ad_m)^2 \a_m^2 - (E_{c,m} + \sqrt{E_{c,q} E_{c,m}}) \ad_m \a_m \\ \nonumber 
& - 2 \sqrt{E_{c,q} E_{c,m}} \ad_q \a_q \ad_m \a_m 
\end{align}
that show up as the Kerr nonlinearity and Lamb shift for the qubit and mediator mode, and a cross-Kerr interaction between the two modes. 

The Hamiltonian in Eq.~\ref{eq:HD} can thus be rewritten into: 
\begin{equation}
\begin{aligned}
    \hat{\mathcal{H}}_D &= \hbar \omega_q \ad_q \a_q - \frac{E_{c,q}}{2} (\ad_q)^2 \a_q^2 + \hbar \omega_m \ad_m \a_m\\
    &- \frac{E_{c,m}}{2} (\ad_m)^2 \a_m^2 + \hbar \chi_{qm} \ad_q \a_q \ad_m \a_m
\end{aligned}
\end{equation}
where $\hbar \omega_q = 4 \sqrt{E_{c,q} E_J} - E_{c,q} - \sqrt{E_{c,q} E_{c,m}}$ is the energy difference between $| 1_q 0_m \rangle$ and $| 0_q 0_m \rangle$, $\hbar \omega_m = 4 \sqrt{E_{c,m} E_J} - E_{c,m} - \sqrt{E_{c,q} E_{c,m}}$ is the energy difference between $| 0_q 1_m \rangle$ and $| 0_q 0_m \rangle$, and $\hbar \chi_{qm} = - 2 \sqrt{E_{c,q} E_{c,m}}$ is the full cross-Kerr shift between the qubit mode and mediator mode. 

\subsection{Mediated dispersive interaction between qubit and resonator}
\label{ssec:mediated_shift}

As illustrated in Fig.~\ref{fig:dimon_device}(d), the symmetry of the dimon circuit along with the coupling capacitances $C_g$ guarantees that the readout resonator is capacitively coupled to only the mediator (dipolar) mode, but not to the qubit (quadrupolar) mode. Using the creation and annihilation operators, the full circuit Hamiltonian can be expressed as:
\begin{align}
    \hat{\mathcal{H}} &= \hat{\mathcal{H}}_D + \omega_r \ad_r \a_r - g_{mr} (\ad_m - \a_m)(\ad_r - \a_r) \\
    &\approx \sum_{k = q,m,r} \omega_k \ad_k \a_k + g_{mr} (\ad_m \a_r + \a_m \ad_r) + \hat{\mathcal{H}}_{\mathrm{nl}} \label{eq:H_coupling}
\end{align}
In the second line we have applied the rotating-wave approximation (RWA). 

We further diagonalize the linear part of the Hamiltonian by a unitary transformation $\hat{U} = \exp{\Lambda (\ad_r \a_m - \a_r \ad_m)}$: 
\begin{equation}
    \hat{U}^{\dagger} \hat{\mathcal{H}} U = \omega_q \ad_q \a_q + \tilde{\omega}_m \ad_m \a_m + \tilde{\omega}_r \ad_r \a_r + 
    \hat{U}^{\dagger} \mathcal{H_{\mathrm{nl}}} \hat{U}
\end{equation}
with $\Lambda = (1/2) \arctan(2g_{mr}/\Delta_{mr})$, where we have denoted the detuning between mediator and resonator as $\Delta_{mr} = \omega_m - \omega_r$. The dressed mode frequencies after the transformation are given by:
\begin{align*}
\tilde{\omega}_m &= \frac{1}{2}\left( \omega_m + \omega_r - \sqrt{\Delta_{mr}^2 + 4 g_{m_r}^2 }  \right)  \approx \omega_m + \frac{g_{mr}^2}{\Delta_{mr}}\\
\tilde{\omega}_r &= \frac{1}{2}\left( \omega_m + \omega_r + \sqrt{\Delta_{mr}^2 + 4 g_{m_r}^2 }  \right) \approx \omega_r - \frac{g_{mr}^2}{\Delta_{mr}}
\end{align*}

Regarding $\hat{U}^{\dagger} \mathcal{H_{\mathrm{nl}}} \hat{U}$ we make the following observations: The first line of Eq.~\ref{eq:H_nl} remains identical under the transformation. The second line of Eq.~\ref{eq:H_nl} is the same as a transmon Hamiltonian, and therefore results in a dispersive shift to the resonator (of $|0_q 1_m\rangle$ versus $|0_q 0_m\rangle$ states): 
\begin{equation}\label{eq:chi_mr}
\chi_{mr} = - 2 \frac{g_{mr}^2 E_{c,m}/\hbar}{\Delta_{mr} (\Delta_{mr} - E_{c,m}/\hbar)}
\end{equation}
We then expand the third line of Eq.~\ref{eq:H_nl} up to 2nd power of $g_{mr}/\Delta_{mr}$:

\begin{equation*}
\begin{split}
\hat{U}^{\dagger} (\hbar \chi_{qm} \ad_q \a_q \ad_m \a_m) \hat{U}
=& \hbar \chi_{qm} \ad_q \a_q \left(\ad_m \a_m + \frac{g_{mr}^2}{\Delta_{mr}^2} \ad_r \a_r  \right. \\ 
-& \left. \frac{g_{mr}}{\Delta_{mr}} (\ad_m \a_r + \a_m \ad_r) \right)
\end{split}
\end{equation*}
Note that the 2nd order perturbation from the last term gives rise to a correction to $\ad_q \a_q \ad_r \a_r$ that is 2nd order in $g_{mr}/\Delta_{mr}$. Accounting for that, the dispersive shift to the resonator (of $|1_q 0_m \rangle$ versus $|0_q 0_m \rangle$ states) has the form: 
\begin{equation}
\chi_{qr} = \frac{g_{mr}^2 \chi_{qm}}{\Delta_{mr}^2} - \frac{(\frac{g_{mr} }{\Delta_{mr}} \chi_{qm})^2}{\Delta_{mr} + \chi_{qm}} = \frac{g_{mr}^2 \chi_{qm}}{\Delta_{mr} (\Delta_{mr} + \chi_{qm})}
\end{equation}

It is worth emphasizing that this \textit{mediated} dispersive interaction between the qubit and resonator does not depend on the qubit-resonator detuning. 
\begin{figure}[b]
    \includegraphics{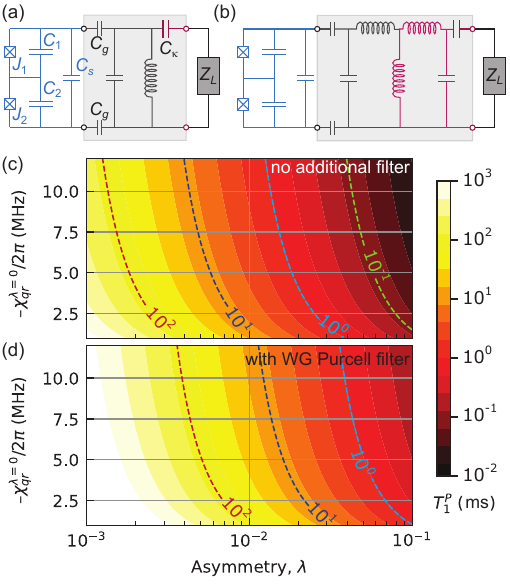} 
    \caption
    [Robustness of Purcell protection]
    {Purcell $T_1$ as a function of junction asymmetry (x-axis) and effective dispersive shift (y-axis). The dissipation seen by the qubit mode is computed using a lumped element circuit simulations with the transmission line modeled as a dissipative load. 
    The weakly coupled port in the experimental device is neglected in this computation.
    } 
    \label{fig:purcell_circuit_sim} 
\end{figure}

\subsection{Purcell decay rate: Circuit model simulation}
\label{ssec:purcell_sim}
The intrinsic Purcell protection of the qubit mode in dimon relies on the geometric symmetry of the device. Typically, the uncertainty in junction fabrication is the main source of asymmetry for our device. We investigate the effect of junction asymmetry, defined as $\lambda= (E_{J1}-E_{J2})/(E_{J1}+E_{J2})$, on the Purcell decay rate of the qubit mode by simulating the circuit in the quantum circuit analyzer tool QuCAT~\cite{qucat}. The anharmonicities and dispersive shifts are computed up to the leading-order non-linear terms in the Hamiltonian.

First, we consider a dimon coupled to a resonator without extra Purcell filtering, as shown in Fig.~\ref{fig:purcell_circuit_sim}(a). We replace the transmission line with a $Z_L = 50\,\Omega$ load coupled to the resonator with a capacitor $C_\kappa$ that sets the external coupling rate of the resonator to $\kappa_r/2\pi = 12$ MHz. First, we construct the circuit with $\lambda = 0$ and choose $E_{J1}=E_{J2}$, $C_1=C_2$ such that the qubit mode frequency and anharmonicity are $\omega_q/2\pi=6.28$ GHz and $\delta_q/2\pi=-0.188$ GHz respectively, motivated by the target values of the experimental device. The dressed readout resonator frequency is kept fixed at $\omega_r/2\pi= 7.5$ GHz. The coupling capacitance $C_g$ and the shunting capacitance $C_s$ are varied to control the mediated qubit-resonator dispersive shift $\chi_{qr}$, while keeping the dressed mediator mode frequency nominally unchanged at $\omega_m/2\pi\sim 4.58$ GHz. Next, we introduce a variable asymmetry in the junction ranging from $\lambda= 10^{-3}$ to $10^{-1}$. As the asymmetry increases, 
the qubit mode inherits a dipole moment and subsequently inherits the losses from the resonator. In Fig.~\ref{fig:purcell_circuit_sim}(c) we plot the Purcell decay time, $T_1^P$ of the qubit mode as a function of the dispersive shift calculated at zero asymmetry, $\chi_{qr}^{\lambda=0}$ and the asymmetry parameter. We observe that even with a few percent of junction asymmetry, $T_1^P$ exceeds few hundred microseconds for $-\chi_{qr}/2\pi=\kappa_r/2\pi = 12$ MHz.

Then, to model the aperture coupled waveguide coupler in our setup we include another filter mode, inductively coupled to the readout resonator (see Fig.~\ref{fig:purcell_circuit_sim}(b)). We set the dressed frequency and external coupling of the filter mode to the experimentally measured values, $\omega_{\rm{PF}}/2\pi = 8.08$ GHz and $\kappa_{\rm{PF}}/2\pi = 0.153$ GHz, and set the shared inductance value such that the effective external coupling of the readout resonator is $\kappa_r/2\pi= 12$ MHz. We perform the same numerical simulation \textemdash We vary the junction asymmetry and plot $T_1^P$ in Fig.~\ref{fig:purcell_circuit_sim}(d). As expected, the additional filter mode provides extra protection against Purcell decay even when there is some residual asymmetry. A few percent of junction asymmetry now results in $T_1^P>1$ ms for $-\chi_{qr}/2\pi=\kappa_r/2\pi = 12$ MHz.

\subsection{Purcell decay rate: Comparison with an equivalent transmon}
\begin{figure}[t]
    \includegraphics{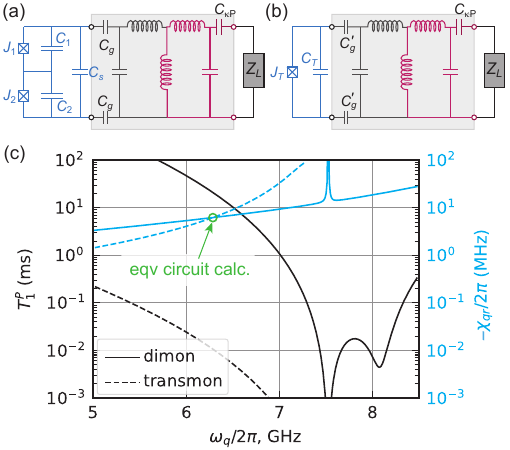} 
    \caption
    [Robustness of Purcell protection]
    {Comparing the the Purcell $T_1$ of a dimon with $\lambda = 1\%$ junction asymmetry with that of an equivalent transmon simulated on a lumped element circuit model. (a) The lumped element circuit used to compute the Purcell $T_1$ of dimon. The circuit includes a readout resonator and a waveguide Purcell filter mode. (b) An equivalent transmon circuit with the identical readout resonator and the waveguide Purcell filter mode. (c) The cyan solid (dashed) curve shows the qubit state dependent dispersive shift of the resonator. The coupling capacitance $C_g'$ of the equivalent transmon circuit is computed to achieve identical dispersive shift, $\chi_{qr}/2\pi = 6.16$ MHz, when the qubit frequency, $\omega_q/2\pi = 6.28$ GHz. The black solid (dashed) curve corresponds to the Purcell $T_1$ of the qubit mode of dimon (transmon), as the frequency of the qubit is tuned from $5$ GHz to $8.5$ GHz by scaling the junction inductance values. The two visible dips in $T_1^P$ of the dimon circuit corresponds to the cases when the qubit mode is on resonance with the readout resonator and the waveguide Purcell filter mode respectively.} 
    \label{fig:purcell_circuit_sim_comp} 
\end{figure}

To contrast the \textit{intrinsic} Purcell protection of dimon (with $1\%$ junction asymmetry) against an ordinary transmon, we simulate both the circuits, including the waveguide filter mode that acts like an external Purcell filter.
We consider identical resonator and the waveguide filter mode parameters for the two cases (kept identical to the previous section). 
We first construct a dimon circuit without any asymmetry and an ``equivalent'' transmon circuit for comparison. We fix $C_1=C_2 = 51.4$ fF and $C_s = 14.4$ fF in the dimon circuit, for this simulation. The junction inductances and the coupling capacitances are chosen to achieve a qubit mode frequency of $\omega_q/2\pi=6.28$ GHz and a qubit state dependent resonator dispersive shift of $\chi_{qr}/2\pi = -6.16$ MHz.
To design an ``equivalent'' transmon we choose the value of the transmon shunting capacitance $C_T$ and junction inductance $J_T$, such that  the transmon anharmonicity is identical to the dimon qubit mode: $\delta_q/2\pi=-0.188$ GHz and the transmon frequency set to $\omega_q/2\pi=6.28$ GHz. 
At the same time, the coupling capacitors for the transmon circuit are tuned such that the resonator dispersive shifts of the transmon is identical to that of the dimon qubit mode, $\chi_{qr}/2\pi = -6.16$ MHz  at this frequency. This point is marked by the small green circle in Fig.~\ref{fig:purcell_circuit_sim_comp}(c).

Next, we introduce $1\%$ asymmetry in the dimon junctions, and scale both the junction inductances with a varying factor to tune the qubit mode frequency of the dimon  and compute $T_1^P$ and the dispersive shift $\chi_{qr}$ as a function of frequency,  as shown by the solid curves in Fig.~\ref{fig:purcell_circuit_sim_comp}(c). We do the same for the junction inductance in the transmon circuit and obtain the corresponding plots for the Purcell $T_1$ and the dispersive shift (dashed curves).  The qubit mode embedded in dimon benefits from three order of magnitude added protection compared to the transmon for a reasonable choice of asymmetry $\lambda= 1\%$.  Note that, changing the junction inductance causes the mediator mode frequency to move, which results in a slow variation of the mediated dispersive shift $\chi_{qr}$ in dimon (solid cyan curve) compared to a more dramatic change in the case of transmon (dashed cyan curve). In dimon, the divergence of $\chi_{qr}$ when $\omega_q=\omega_r$ is due to the asymmetry and resulting linear hybridization between the qubit mode and the resonator.

\section{Experimental device and measured parameters}
In this section we discuss the device parameters, experimentally measured thermal population, preparation and readout of the non-computational ``leakage'' states presented in the main text Fig.~1, and Purcell decay rate of the qubit mode.

\begin{table}[b!]
\centering
 \begin{tabular}{|l | c | c|} 
 \hline
 Entity & Qubit & Mediator \\ [0.5ex] 
 \hline
 Frequency ($\omega/2\pi$) (GHz) & $6.271$& $4.633$ \\ 
 Anharmonicity ($\delta/2\pi$) (GHz) & $-0.183$& $-0.101$  \\ 
 Dispersive interaction ($\chi/2\pi$) (MHz) & $-6.4$ & $-8.5^*$  \\ 
 Relaxation time ($T_1$) ($\mu$s) & $50.$ & $71$  \\ 
 Ramsey time ($T_2^R$) ($\mu$s) & $36$ & $18$ \\ 
 Hahn echo time ($T_2^E$) ($\mu$s) & $42$ & $61$ \\
 \hline
 \end{tabular}
 \caption{Device parameters and coherence of the two modes of dimon. 
 $^*$Mediator mode dispersive shift is measured in a different cool-down with identical set up.}
 \label{table:device_parameters}
\end{table}

\subsection{Device details} 
\label{appendix:device_params}
The device is fabricated on Heat Exchanger Method (HEM) sapphire with thin film aluminum capacitor pads and the Al-AlOx-Al junctions are formed by bridge-free technique. Both the readout cavity and the waveguide are made of 6061 aluminum alloy.

We characterize the eigen-energies of the composite Hilbert space of the dimon by preparing it in the states $|0_q0_m\rangle$, $|1_q0_m\rangle$, $|0_q1_m\rangle$, $|2_q0_m\rangle$, $|1_q1_m\rangle$, and $|0_q2_m\rangle$, where the subscripts $\{q,m\}$ represent the mode which is excited.  
The relevant frequencies, anharmonicities and the total resonator dispersive shift for the two modes of dimon are listed in Table~\ref{table:device_parameters} along with the measured coherence of the two modes of the dimon.
The cross-Kerr interaction between the qubit and the mediator mode is measured to be $\chi_{qm}/2\pi = -272$ MHz. The readout resonator has a dressed frequency $\omega_r/2\pi=7.515$ GHz when the dimon is in the ground state.

\subsection{Thermal population}
\label{ssec:thermal}

\begin{figure}[t!]
\includegraphics{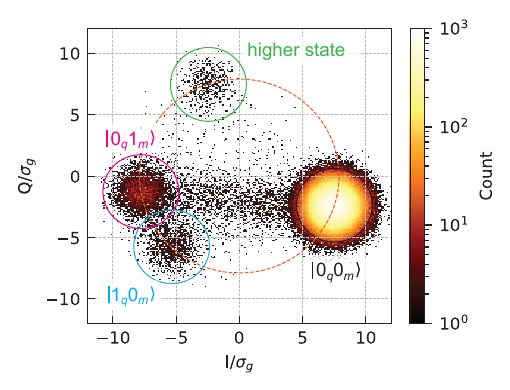}
\caption
[Thermal population]
{
Thermal population of dimon measured through continuous wave reflection spectroscopy of the readout resonator using a VNA.
}
\label{fig:thermal_pop} 
\end{figure}

We perform a CW measurement using a vector network analyzer (VNA) to measure the thermal population in the first excited state of the qubit mode, $|1_q0_m\rangle$ and the mediator mode, $|0_q1_m\rangle$ of the dimon. 
We measure a population of 0.26\% and 1.30\% respectively for the $|1_q0_m\rangle$, and, $|0_q1_m\rangle$ states from the phase space histograms (plotted in Fig.~\ref{fig:thermal_pop}).

Assuming the system is in thermal equilibrium, we estimate an effective temperature of the device by computing the thermal population in the higher excited states of the dimon using the expression~\cite{jin_2015}:
\begin{equation}
    P_{|i\rangle}=\frac{\exp(-E_i/k_BT)}{\sum_i{\exp(-E_i/k_BT)}}
\end{equation}
We consider the first four levels $|0_q0_m\rangle$, $|0_q1_m\rangle$, $|1_q0_m\rangle$, and, $|0_q2_m\rangle$, beyond which the thermal population should be negligible. The measured higher state populations are consistent with a temperature $T = 51$ mK for our device. 
This measurement also highlights the need for a \textit{pre-selection readout} in our experiments, bringing down the thermal population well below $0.01\%$.   

\subsection{Preparation and readout of non-computational states}
We use a series of $\pi$-pulses at relevant transition frequencies to prepare the dimon in three known ``leakage'' states to demonstrate how these higher energy states may overlap with the two computational states of the dimon in Fig.~1 of the main text. These prepared states, $|l_1\rangle$, $|l_2\rangle$ and $|l_3\rangle$ correspond to $|2_q0_m\rangle$, $|1_q1_m\rangle$, and $|1_q2_m\rangle$ respectively. The subscripts $\{q,m\}$ represent the mode which is populated. Note that these states are only to demonstrate of how the readout may fail to distinguish these states from the $|g\rangle$ and $|e\rangle$ states and these prepared ``leakage'' states may not be the states the qubit leaks into during the readout operation.

To perform a readout of these prepared states, we operate the quantum-limited amplifier in the phase-preserving mode and integrate for $400$ ns to obtain the readout histograms presented in Fig.~1. The readout power is adjusted to separate the mean of the $|g\rangle$ and $|e\rangle$ histograms by more than $3(\bar{\sigma}_g+\bar{\sigma}_e)$, where $\bar{\sigma}_g$ and $\bar{\sigma}_e$ are the standard deviations of the projected $|g\rangle$ and $|e\rangle$ histograms, respectively.

\subsection{Measurement of Purcell decay rate} 
\label{appendix:rabi_rate}
\begin{figure}[t!]
\includegraphics{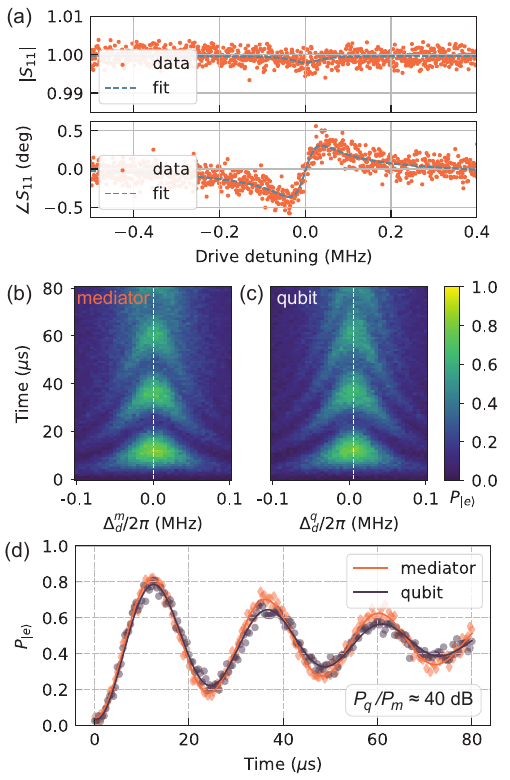}
\caption
[Estimation of Purcell protection from Rabi rate]
{(a) The reflection spectrum of the mediator mode from direct continuous wave scattering and its fit to a model for complex reflection coefficient with a saturable mode. (b-c) Rabi oscillations on the two modes of dimon as a function of drive detuning. (d) Near identical Rabi rates, $\Omega_m/2\pi = 41.7$ kHz and $\Omega_q/2\pi = 41.2$ kHz obtained by driving with a power ratio $P'_q/P'_m\sim43$ dB measured at the room temperature. Taking into account of the variation of input attenuation of the cryogenic setup at the two different frequencies, this translates to $40$ dB power difference at the strongly couple port of the resonator.
   }
\label{fig:rabi_rate} 
\end{figure}

Since the Purcell decay rate of the qubit mode is expected to be very small, it is difficult to directly measure it in experiments. Hence, we take a two-step approach to estimate $T_1^P$ of the qubit mode. We first measure the Purcell decay rate of the mediator mode which directly couples to the TE101 mode of the readout cavity and can spontaneously radiate into the strongly coupled port. We perform a direct emission spectroscopy~\cite{mirhosseini_2019_atom_cavity, characterizing_direct_mw, sunada_2022} to measure the Purcell decay rate of the mediator mode. We apply a weak saturation tone on the mediator mode at its resonant frequency and fit the reflected quadrature signal to the expression (See Fig.~\ref{fig:rabi_rate}(a)):
\begin{equation}
    S_{11}(\omega_d)=1-\frac{\tilde{\Gamma}_1^P}{\Gamma_2}\frac{1-i(\omega_m-\omega_d)/\Gamma_2}{1+s+(\omega_m-\omega_d)^2/\Gamma_2^2}
\end{equation}
where $s$ is a fit parameter reflecting the population saturation of the mode under the CW drive,
\begin{equation}
    s = \frac{\Omega^2}{(\Gamma_{\uparrow}+\Gamma_{\downarrow})\Gamma_2}
\end{equation}
and, finally,
\begin{equation}
    T_1^P=\frac{1}{\Gamma_1^P} = \frac{1-r_{\rm{th}}}{1+r_{\rm{th}}}\frac{1}{\tilde{\Gamma}_1^P},
\end{equation}
where, $r_{\rm{th}}\equiv P_{0_q1_m}/P_{0_q0_m}$ is the thermal excitation ratio measured independently (and reported in Sec.~\ref{ssec:thermal}). From the fit, we find a Purcell decay time for the mediator mode, $T_{1,m}^P = 0.34$ ms. 

To compute the Purcell decay rate of the qubit mode we compare the Rabi oscillation rates of the mediator and the qubit mode, when driven from the strongly coupled port.
The Rabi Rate $\Omega$ of a mode for a certain drive power $P$ applied on a given port is directly related to the radiation decay of that mode through that port:
\begin{equation}
    \Gamma_1^P(\omega_d) = \frac{\Omega^2\hbar\omega_d}{4P}
    \label{eq:rabi_to_purcell}
\end{equation}
We measure a power ratio of $P'_q/P'_m\approx43$ dB, at the input of the dilution fridge ($300$ K), to achieve near identical Rabi rates of  $\Omega_m/2\pi = 41.7$ kHz and $\Omega_q/2\pi = 41.2$ kHz on the mediator and qubit mode respectively. From an independent calibration of input line attenuation, this translates to a power ratio of $P_q/P_m\approx40$ dB at the strongly-coupled port of the resonator. We use this calibration to estimate a Purcell $T_{1,q}^P = 2.6$ seconds for the qubit mode.
Note that the measured Purcell $T_1$ in experiments exceeds the value predicted from the lumped circuit simulation by two orders of magnitude. This enhancement can be attributed to the multi-mode effect of the 3D cavities~\cite{houck_purcell}. In our simulation we only kept the two lowest frequency modes of the distributed 3D architecture containing the readout cavity and the waveguide Purcell filter.

Finally, to compute the Purcell decay rate through the weakly coupled port, we compare the Rabi rate on the qubit mode through the strong port and the weak port. Using~\ref{eq:rabi_to_purcell} and the calibration of input attenuation of the drive line and the readout line, the Purcell decay rate through the weakly coupled port is estimated to be $0.4$ seconds.

\section{Amplification chain and measurement efficiency}

\begin{figure}[t]
\includegraphics{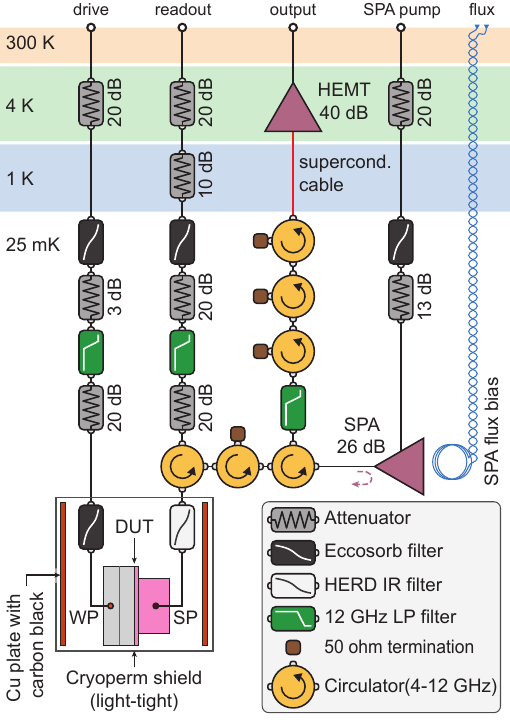} 
\caption
[Experimental setup]
{Cryogenic experimental setup at different stages of the dilution fridge. 
}
\label{fig:setup_wiring} 
\end{figure}
Our experiments are conducted in an Oxford Triton\textsuperscript{TM} dilution refrigerator with a base temperature reaching $T_{\rm{MXC}}=28$ mK. A schematic of the cryogenic experimental set-up is shown in Fig.~\ref{fig:setup_wiring}. The 3D cavity along with the waveguide coupler is placed inside a hollow copper cylinder lined with carbon black and sealed inside a light-tight Cryoperm magnetic shield. The copper cylinder thermalizes with the mixing chamber of the dilution fridge. 
To protect the device from infrared radiation from the transmission lines, we employ an Eccosorb filter at the weakly coupled port and a high-energy radiation drain (HERD)~\cite{herd_2023} filter at the strongly coupled port. 
The HERD filter has negligible losses at readout frequency, and therefore causes minimal efficiency degradation to the output line.
Additionally, we use a low insertion loss triple-junction cryogenic circulator array ($S_{21}\sim0.5$ dB total) to further minimize losses before the SNAIL parametric amplifier (SPA). We operate the SPA in the phase-sensitive mode and calibrate it to produce $26$ dB gain to suppress the added noise after the SPA. 
\begin{figure}[t]
    \centering
    \includegraphics{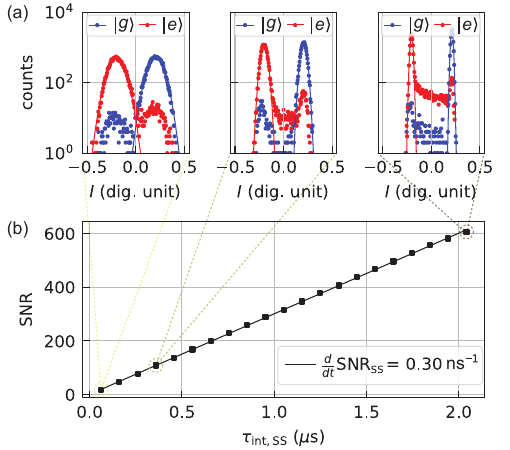}
    \caption{Steady state SNR vs time 
    with $\bar{n}_r = 2.8$.(a) Integrated quadrature histograms consisting of $19\,000$ shots, projected on the I quadrature. Three such sets obtained with different integration times are shown. A Gaussian fit is used on each of them to extract the discriminability and SNR of the signal. (b) Plotted readout SNR as a function of integration time and a linear fit to extract the readout efficiency.}
    \label{fig:efficiency}
\end{figure}

The measurement efficiency $\eta$ of the output chain is measured using the expression for the steady state SNR of the integrated signal.
\begin{equation}
    {\rm{SNR}}_{\rm{SS}} = \eta\bar{n}_r\kappa_{\rm{ext}}\tau_{\rm{int}}\frac{8\chi_{qr}^2}{\chi_{qr}^2+\kappa_r^2}
\end{equation}
The external coupling rate $\kappa_{\rm{ext}}$ is power independent and is calculated from a circle fit on the quadrature signal from a low power reflection measurement. The total readout resonator line-width $\kappa_r = \kappa_{\rm{ext}} + \kappa_{\rm{int}}$ and the intra-resonator photon number are simultaneously measured from an independent experiment explained in Sec.~\ref{appendix: readout_kerr}. $\kappa_{\rm{int}}$ is the effective internal loss rates of the readout resonator. The readout resonator dispersive shift $\chi_{qr}$ is calculated from the resonator response (phase-roll) when the qubit is prepared in $|g\rangle$ and $|e\rangle$ states. ${\rm{SNR}}_{\rm{SS}}(\tau_{\rm{int}})$ is measured from the integrated quadrature histograms in the steady state for different integration times $(\tau_{\rm{int}})$. 
We use a linear fit to extract a slope:
\begin{equation}
    \frac{\mathrm{d} {\rm{SNR}}_{\rm{SS}}}{\mathrm{d} \tau_{\rm{int}} }= \eta\bar{n}_r\kappa_{\rm{ext}}\frac{8\chi_{qr}^2}{\chi_{qr}^2+\kappa_r^2} = 0.30 \, \mathrm{ns}^{-1}
\end{equation}
The readout efficiency of the output chain (with $26$ dB gain from SPA) is measured to be 
$\eta = 0.79 \pm 0.18$
from this relation, with the uncertainty dominantly arising from the fitting uncertainty of $\kappa_{\rm{ext}}/2\pi = 11.6 \pm 1.4$ MHz.

\section{Readout of the qubit mode}
We perform a dispersive readout of the qubit mode in reflection. 
The readout resonator is driven from the strongly coupled port at frequency $\omega_d \approx \omega_r+\chi_{qr}/2$ to maximize the differential phase shift, where $\omega_r$ is the dressed resonator  frequency when the dimon is in the ground state.
The reflected wave is integrated for a time $\tau_{\rm{int}}$ to obtain a demodulated quadrature signal $S(\tau_{\rm{int}})=\sqrt{\kappa_r}\int_0^{\tau_{\rm{int}}}a_{\rm{out}}(t)\mathcal{K}(t){\rm{d}}t$. $\mathcal{K}(t)$ is an experimentally obtained demodulation envelope~\cite{optimal_readout}, proportional to $|\langle a_{\rm{out}}^g(t)-a_{\rm{out}}^e(t)\rangle|$, where $a_{\rm{out}}^{g,e}(t)$ is the instantaneous output wave amplitude corresponding to the qubit being in $|g\rangle$ (ground state) or $|e\rangle$ (excited state).
The optimization of the readout pulse involves the photon number calibration, finding the optimum readout frequency and finally tuning the amplitudes of the piece-wise constant pulse that expedites the evacuation of the readout resonator at the end of the readout. We also show that the self-Kerr of the readout resonator is non-negligible and consequently the pulse-shape needs to be independently optimized at each power.

\subsection{Resonator non-linearity and photon number calibration} 
\label{appendix: readout_kerr}
In our device, to achieve a large dispersive shift, we strongly hybridize the mediator mode with the readout resonator. This causes the readout resonator to inherit some non-linearity from the participation of the Josephson junctions in the readout mode. 
Although, during the readout we use a relatively small photon number compared to the critical photon number to bifurcate the readout resonator ($\bar{n}_r \ll -\kappa_r/3\sqrt{3}K$), the self-Kerr shift of the readout frequency is non-negligible.
This has a rather small effect on the SNR, but crucially modifies the choice of drive detuning in readout pulse shaping. 
Furthermore, owing to the inherited self-Kerr, the \textit{average photon number} in the readout resonator exhibits a non-linear reliance on the input drive power, necessitating a careful calibration of the average photon number.

To simultaneously calibrate the average photon-number and measure the Kerr non-linearity of the readout resonator, we assume a Duffing oscillator model for the readout resonator~\cite{non_linear_rdt_res}, where the leading non-linear term is given by $K$. 
\begin{equation}
    \dot{\alpha}(t) = -\left(i\Delta_d+iK|\alpha(t)|^2+\frac{\kappa_r}{2}\right)\alpha(t) +A(t).
    \label{eq:duffing_ode}
\end{equation}
Here, $\Delta_d$ is the detuning of the drive from the resonance (measured at low power limit), $\alpha(t)$ is the instantaneous displacement of the oscillator in the phase space
and $A(t) = \sqrt{\kappa_r}\bar{a}_{in}(t)$, where $\bar{a}_{in}(t)$ is the instantaneous amplitude of the input photon field. 
Since we do not have an accurate measure of drive amplitude at the readout port, we assume a complex scaling factor $k_{\rm{sca}}$ (to be determined from the fit) such that, $A(t)=k_{\rm{sca}} a_{\rm{DAC}}(t)$. 
We also assume that $k_{\rm{sca}}$ is frequency independent for the $30$ MHz range of drive detuning $\Delta_d$ used in the experiment.

\begin{figure}
\centering
\includegraphics[width = 0.48\textwidth]{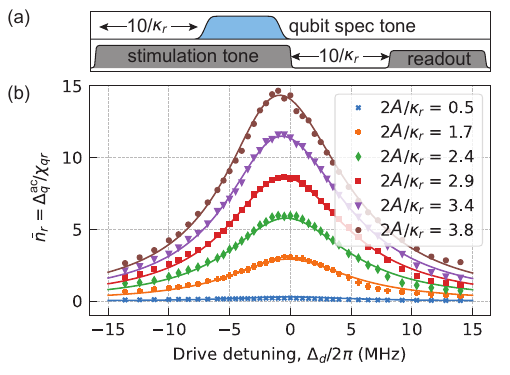}
\caption
{
Measurement of the acquired non-linearity of the readout resonator.
(a) The pulse sequence to measure average photon number in the readout resonator $\bar{n}_r$, as a function of the detuning and the amplitude of the applied readout tone.
(b) The estimated intra-resonator photon number, measured from the ac Stark shift on the qubit
is fitted to the Kerr oscillator model. From this fit, we simultaneously estimate the Kerr non-linearity, resonator line-width and the input line attenuation.
}
\label{fig:readout_kerr} 
\end{figure}

The classical steady-state response of such a Kerr oscillator, to the leading order of non-linearity is given by~\cite{non_linear_rdt_res, yurke_2006}:
\begin{equation}    \left(i\Delta_d+iK|\alpha_{\rm{ss}}|^2+\frac{\kappa_r}{2}\right)\alpha_{\rm{ss}} =A
    \label{eq:duffing_steady_state}
\end{equation}
The steady-state amplitude of the resonator is related to the intra-resonator photon number by $|\alpha_{\rm{ss}}|^2 = \bar{n}_r$. The latter can be quantified in terms of the ac Stark shift on the qubit and the qubit-resonator dispersive shift $\chi_{qr}$. 
Experimentally, we perform a qubit spectroscopy under a resonator stimulation tone with varying $\Delta_d$ and $a_{\rm{DAC}}=A/k_{\rm{sca}}$ to record the ac Stark shift. 
The resonator stimulation tone is turned on $10/\kappa_r$ prior to the low power qubit probe tone around the qubit frequency, making sure the resonator has reached the steady-state. We then turn off the stimulation tone and let the resonator ring-down for a duration $10/\kappa_r$, followed by a qubit readout. 
In Fig.~\ref{fig:readout_kerr}(b) we plot the intra-resonator photon number (at steady-state) inferred from the ac Stark shift as a function of drive detuning $\Delta_d$ for different DAC amplitudes $a_{\rm{DAC}}$.
We fit the dependence with the Kerr-oscillator response given by Eq.~\ref{eq:duffing_steady_state} with three fit parameters, $K$, $\kappa_r$ and, $|k_{\rm{sca}}|$, ($\{\kappa_r, K\}\in \mathbb{R}, k_{\rm{sca}}\in\mathbb{C}$).
\begin{equation}
    \left[\left(\Delta_d+K\bar{n}_r\right)^2+\frac{\kappa_r^2}{4}\right]\bar{n}_r =|k_{\rm{sca}}|^2 |a_{\rm{DAC}}|^2
    \label{eq:steady_state_stark_shift}
\end{equation}
We have absorbed the phase of the displacement $\alpha$ on the LHS of the equation in the complex scaling factor $k_{\rm{sca}}$. We obtain $\kappa_r/2\pi = 12$ MHz, $|k_{\rm{sca}}| = 162$, and $K/2\pi= -60.$ kHz. 
Note that the estimated $\kappa_r$ from this experiment differs by about $3.4\%$ from the one extracted from the direct reflection measurement of the resonator performed with a small steady state photon number, $\bar{n}_r \sim 2.8$. This is due to parametric excitation of the dimon to leakage states at higher probe power, causing an increase of effective line-width of the resonator.

\subsection{Readout pulse shaping}  

To speed up the readout resonator ramp-up and ramp-down, we experimentally calibrate a four-stage readout pulse as described in the main text. 
Our goal is to maximally separate the two resonator trajectories $\alpha_g(t)$ and $\alpha_e(t)$ corresponding to the $|g\rangle$ and $|e\rangle$ states of the qubit respectively, under the constraint of a maximum photon number and an unconditional reset of the readout resonator at the end of the pulse.
Each section of the resonator control pulse intends to drive the resonator from one steady state to another. 
The initial high-amplitude response rapidly drives the resonator towards a large displaced state. But as soon as we reach the desired maximum photon number ($\bar{n}_{\rm{max}}$) for the readout, we release the resonator to a smaller drive amplitude, letting it evolve towards the new steady state, maintaining the photon number below $\bar{n}_{\rm{max}}$. The third stage speeds up the resonator once again, but this time in the opposite direction till we find a suitable point in phase space for a sharp ``U-turn'' and converge at the origin with the help of another high-amplitude segment. The third and fourth segment of the pulse causes a rapid driven ramp-down. 

To experimentally optimize the shape of the four-stage pulse, we constrain the maximum photon number $\bar{n}_{\rm{max}}$ during the evolution and the total pulse duration. We vary the relative length and the amplitude of each section to minimize the ratio of undesired measurement after the pulse ends to the desired measurement during the pulse. The measurement is quantified in terms of the integrated SNR over a period. Putting in a compact form, the shaped pulse corresponds to the optimization problem:
\begin{equation}
\begin{split}
\minimize_{\{A_i, l_i, \Delta_0\}\in \mathbb{R}}  \vartheta(A_i, l_i, \Delta_0),\quad \ni|\alpha_g|^2, |\alpha_e|^2 < \bar{n}_{\rm{max}}&\\ {\rm{and}} \: A_i\neq0, {\rm{and}} \:\sum_i{l_i} = \tau_{\rm{ro}}&\\
   \vartheta(A_i, l_i, \Delta_0)=\frac{\int_{\tau_{\rm{ro}}}^{\infty}|\alpha_g(t)-\alpha_e(t)|^2dt}{\int_{0}^{\tau_{\rm{ro}}}|\alpha_g(t)-\alpha_e(t)|^2dt},&
\end{split}
\end{equation}
By allowing the pulse detuning ($\Delta_0$) to be a free parameter, we can compensate to the leading order the asymmetry due to the resonator non-linearity for small $K$.

\subsection{Delay caused by the SPA}
\begin{figure}[t!]
    \includegraphics{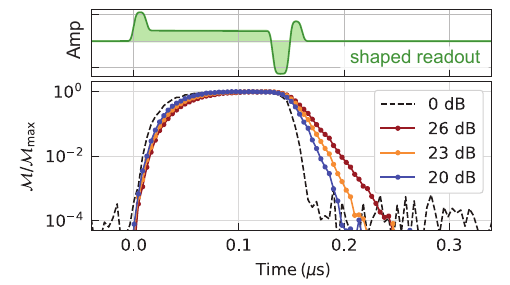}
    \caption
    [SPA slowing down the readout wave-packet]
    {Slowing down of the reflected wave-packet containing the state information of the qubit after passing through the SPA. The normalized reflected readout signal is plotted against time when the SPA is biased to produce different gain. A higher gain, makes the bandwidth narrower causing a more significant delay. 
    }
    \label{fig:pulse_slow_down_SPA} 
\end{figure}

At the chosen operation point that provides $26$ dB gain, the SNAIL parametric amplifier (SPA) in our setup has a relatively smaller bandwidth, $8.3$ MHz, compared to the readout resonator. This causes slowing down of the readout wave-packet emitted from the resonator. To calibrate and reliably assess the performance of the shaped pulse, therefore, we turn of the SPA and perform $100\,000$ shots of measurements to achieve sufficient SNR. We show this with the black dashed curve in Fig.~\ref{fig:pulse_slow_down_SPA}, by plotting the normalized measurement rate ($\mathcal{M}/\mathcal{M}_{\rm{max}}$) vs time for the entire duration of the pulse and afterwards. We then tune the SPA to different gain settings ranging from $20$ dB to $26$ dB and show that as the gain increases, the SPA bandwidth is reduced, thus worsening the slowdown (shown with blue, yellow and dark red traces, respectively, averaged over $20\,000$ shots). Therefore, to capture the entire wave-packet and acquire the complete information, one needs to integrate longer than the readout duration itself. This highlights the need for broadband parametric amplifiers in the readout amplification chain for our future experiments to carry out fast measurements.

\subsection{Review of other dispersive readouts and Purcell filters} 
\begin{figure}[t]
    \centering
    \includegraphics[width=0.48\textwidth]{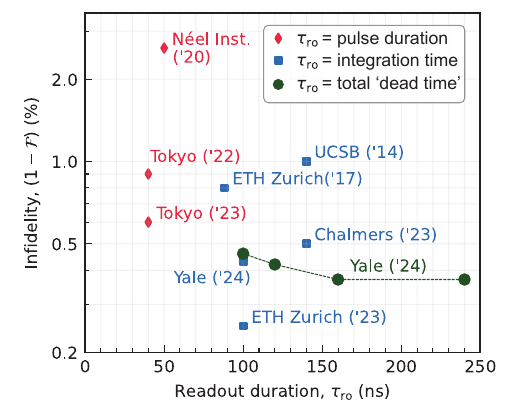}
    \caption{State of the art readout fidelity and readout duration. Different colors and symbols represent different definitions used to report the ``readout duration''. 
    The red diamonds represent the references where it refers to the time-span of the applied readout pulse, and the blue squares represent those where it is defined as the integration time. In this work, we have defined the readout duration that includes the resonator ramp-up and ramp-down times and the corresponding data is shown by dark green circles. For comparison, we have also reported our result in terms of integration time, with one data point at $100$ ns. Note that, Chalmers ('23)~\cite{chen_2022} uses excited state promotion techniques. }
    \label{fig:fidelity_refs}
\end{figure}

We compare the readout performance and the device properties in our experiments with other related works reported earlier. In Table~\ref{table:ref_pars_coherence} we list the readout parameters such as the external coupling of the readout resonator ($\kappa_r$), dispersive interaction between the qubit and the readout resonator ($\chi_{qr}$) and the \textit{undriven} relaxation time ($T_1$) of the qubit in previous works.
In Fig.~\ref{fig:fidelity_refs}, we list the reported readout duration and the readout infidelity in these works. However, for the lack of unified convention to define the readout duration, we use different color and markers to label the reported numbers. The red diamonds, represent the experiments stating readout pulse duration and the blue squares refer to the experiments where the readout duration is defined with respect to the integration time. Here, we introduce a stricter definition (dark green circles, corresponding to the four readout pulses we compared in the main text) to account for the resonator ramp-up and ramp-down times within the readout duration.

\begin{table}[t!]
\centering
\begin{tabular}{|l|*{3}{c|}}\hline
\backslashbox{References}{Parameters}
&\makebox[3.5em]{\begin{tabular}{@{}c@{}}$\kappa_r/2\pi$ \\ (MHz)\end{tabular}}&\makebox[3.5em]{\begin{tabular}{@{}c@{}}$\chi_{qr}/2\pi$ \\ (MHz)\end{tabular}}&\makebox[4em]{\begin{tabular}{@{}c@{}}Undriven \\ $T_1$ ($\mu$s)\end{tabular}}\\
[1ex] 
\hline
 USCB $^\dagger$, 2014~\cite{jeffrey_2014} \quad  & 4.3 & N.A. & 11 \\ 
 TU Delft, 2016~\cite{bultink_2016} \quad  & 0.62 & -2.6 & 25 \\ 
 ETH, 2017~\cite{walter_2017} \quad  & 37.5 & -15.8 & 7.6 \\
 Institut Néel, 2020 \cite{dassonneville_2020} \quad  & 12.1 & -9.0 & 3.3 \\ 
 Tokyo, 2022~\cite{sunada_2022} \quad  & 45.7 & -6.9 & 17 \\ 
 Chalmers, 2023~\cite{chen_2022} \quad & 11 & -12.6 & 6.2\\ 
 Tokyo, 2023 \cite{sunada_2023} \quad & 11.8 & -11.8 & 17 \\
 ETH, 2023~\cite{swiadek_2023} \quad & 25.0 & -6.3 & 30.\\ 
 \textbf{This work, 2024} \quad  & \textbf{11.6} & \textbf{-6.4} & \textbf{50.} \\
 [1ex] 
 \hline
\end{tabular}
 \caption{Cavity line-width, dispersive interaction and idle device T1, when there is no photons in the readout cavity, for the references in Fig.~\ref{fig:fidelity_refs}. $^\dagger$Information about dispersive interaction strength $\chi$ is not provided in the reference.}
 \label{table:ref_pars_coherence}
\end{table}

\section{Error channel in the 
presence of readout-induced leakage}
\begin{figure}
\centering
\includegraphics{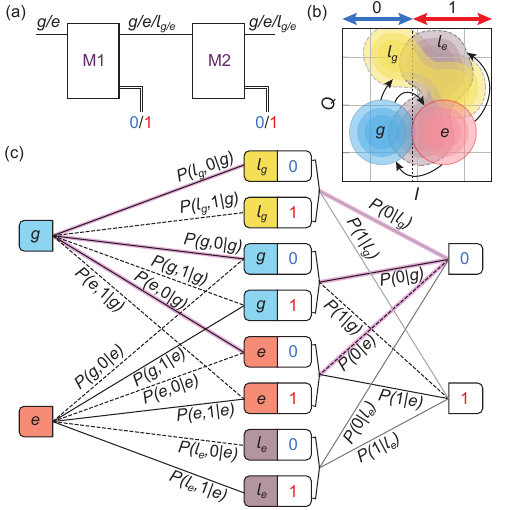}
\caption
{Readout error model. (a) Modeling the binary readout channel for the repeatability experiment consisting of two back-to-back measurements (M1 and M2). (b) A cartoon of readout signal on the IQ plane before the binary projection, illustrating the transition errors either into the wrong computational state or into the leakage states from $g$ and $e$ states (shown in yellow and gray, respectively). The vertical dashed line denotes the demarcation threshold for the binary readout outcome. (c) The error model for the repeatability experiment. The model is characterized by the probabilities of obtaining a post-measurement state and a readout outcome, given an initial state of the qubit. The dashed lines represent the errors that are captured in the \textit{readout fidelity} metric. To compute the repeatability metric, we decompose the probability that two successive readout operations produce ``$0$'' outcomes when the qubit is initialized in the $g$ state. This decomposition is shown by the line segments highlighted in purple. The three mutually exclusive paths (depending on the qubit states after the first measurement) add together. This decomposition shows how transition errors, in particular leakage errors can be overlooked in the \textit{repeatability metric}, when probabilities such as $P(0|l_g)$ are of the order of unity.
}
\label{fig:repeatability_channel} 
\end{figure}

The binary readout channel can be characterized as a process that acts on an input quantum state $\{g,e\}$, produces a classical bit $\{0,1\}$ that corresponds to the readout outcome and leaves the quantum system in a ``post-measurement'' state $\{g,e, l_{g/e}\}$, as shown in Fig.~\ref{fig:repeatability_channel}(a). 
Here, $l_{g/e}$ is a mixed state consisting of all the leakage states into which the qubit is excited by the readout operation. In general, the leakage states for different qubit initializations $\{g,e\}$ could be different and their readout signal may or may not overlap in the IQ plane, as illustrated in Fig.~\ref{fig:repeatability_channel}(b).We also show the demarcation threshold in the IQ plane via the dashed vertical line. Any readout outcome to the left (right) of the demarcation threshold is assigned as ``$0(1)$''.

Therefore, the binary readout error channel can be modeled via the probabilities of producing the readout result `$0$' or `$1$' for the two input states `$g$' and `$e$' and for all possible post-measurement states, $\{g, e, l_g, l_e\}$~\cite{msp_riste_2012}. 
Each of the line segments in Fig.~\ref{fig:repeatability_channel}(c) corresponds to one such probability. For example, $P(0, l_e|e)$ corresponds to the probability that measuring an input state $e$ produces the readout outcome `$0$' and the post-measurement state is $l_e$. 
Experimentally, the ``post-measurement'' state of the first readout is typically probed by a second readout, which can be modeled using another set of probabilities. We can then define the standard metrics to characterize the binary readout in terms of these probabilities. 

The \textit{readout fidelity} is defined as: $\mathcal{F} := [\mathcal{F}_g + \mathcal{F}_e]/2$. We can break it further into:
\begin{equation}
\begin{split}
    \mathcal{F}_g = P(0|g) = P(g,0|g) +P(e,0|g)+P(l_g,0|g),\\
    \mathcal{F}_e = P(1|e)= P(g,1|e) +P(e,1|e)+P(l_e,1|e)
\end{split}
\end{equation}
Note that this metric does not care about the post-measurement state. 

We define the \textit{QND measurement fidelity} as:
\begin{equation}
    \begin{split}
        \mathcal{F}^{\rm{QND}}_g :=& P(g, 0|g),\\
        \mathcal{F}^{\rm{QND}}_e :=& P(e, 1|e)\\
         \mathcal{F}^{\rm{QND}}:=&  \frac{1}{2}\left[\mathcal{F}^{\rm{QND}}_g+ \mathcal{F}^{\rm{QND}}_e\right]
    \end{split}
\end{equation}
for the qubit initializations $|g\rangle$ and $|e\rangle$, respectively.
We reserve the symbol $\mathcal{Q}$ to denote ``\textit{QNDness}''\cite{lupacscu_2007} that quantifies the degree to which the readout leaves the qubit state unaffected, regardless of the readout outcome,  $\mathcal{Q}= [ \mathcal{Q}_g+\mathcal{Q}_e]/2 $. 
\begin{equation}
\begin{split}
    \mathcal{Q}_g= P(g,0|g)+P(g,1|g),\\
    \mathcal{Q}_e= P(e,0|e)+P(e,1|e)  
\end{split}
\end{equation}
$\mathcal{F}^{\rm{QND}}_g$ and $\mathcal{Q}_g$ thus differ only by the readout SNR error $P(g,1|g)$.

Finally, we express repeatability, the quantity that is used as a proxy for ``QNDness'' as: $\mathcal{R}= [\mathcal{R}_g +\mathcal{R}_e ]/2$, where $\mathcal{R}_g := P(0_{\rm{M2}}|0_{\rm{M1}}, g)$ and $\mathcal{R}_e := P(1_{\rm{M2}}|1_{\rm{M1}}, e)$. The notation, $P(0_{\rm{M2}}|0_{\rm{M1}}, g)$ denotes the probability that the first measurement produces ``$0$'' and the second measurement produces ``$0$'' given the qubit is prepared in ``$g$'' prior to the first readout. Using the conditional probability identities we can write:
\begin{equation}
\mathcal{R}_g = P(0_{\rm{M2}}|0_{\rm{M1}}, g) = \frac{P(0_{\rm{M2}}, 0_{\rm{M1}}|g)}{P(0|g)} 
\label{eq:repeatability_def}
\end{equation}
The repeatability can then be decomposed in terms of the probabilities already introduced in the readout error channel. 
As illustrated in Fig.~\ref{fig:repeatability_channel}(c) with the highlighted line segments, the numerator can be decomposed into three terms:
\begin{equation}
    \begin{split}
        P(0_{\rm{M2}},0_{\rm{M1}}| g) &= P(g,0|g) P(0|g) \\
        &+ P(e,0|g) P(0|e)\\
        &+ P(l_g,0|g) P(0|l_g) 
    \end{split}
\end{equation}
We can use this relation to connect the repeatability and QND measurement fidelity:
\begin{equation}
    \begin{split}
        \mathcal{R}_g =&\mathcal{F}^{\rm{QND}}_g
        + P(e,0|g) \frac{P(0|e)}{\mathcal{F}_g} + P(l_g,0|g) \frac{P(0|l_g)}{\mathcal{F}_g} \\
        =&\mathcal{F}^{\rm{QND}}_g + \Xi_g
    \end{split}
\end{equation}
The symbol $\Xi_{g/e}$ is always a positive quantity and denotes the probability of overlooking a transition error from the $g/e$ state by two successive readout operations.
\begin{equation}
    \begin{split}
        \Xi_g= P(e,0|g) \frac{P(0|e)}{\mathcal{F}_g} + P(l_g,0|g) \frac{P(0|l_g)}{\mathcal{F}_g} \\
        \Xi_e= P(g,1|e) \frac{P(1|e)}{\mathcal{F}_e} + P(l_e,1|e) \frac{P(1|l_e)}{\mathcal{F}_e}
    \end{split}
\end{equation}

Therefore, repeatability $\mathcal{R}$ always overestimates the term, QND measurement fidelity, $\mathcal{F}^{\rm{QND}}$ defined this way.
\begin{equation}
\mathcal{R}=\mathcal{F}^{\rm{QND}} + \Xi ; \quad \Xi := \frac{\Xi_g+\Xi_e}{2}
\end{equation}
We can also write down the relationship between repeatability $\mathcal{R}$ and QNDness $\mathcal{Q}$ as:
\begin{equation}
    \begin{split}
        \mathcal{R} =&\mathcal{Q} - \frac{P(g,1|g)+P(e,0|e)}{2} + \Xi
    \end{split}
\end{equation}
Thus, repeatability $\mathcal{R}$ can underestimate or overestimate QNDness, $\mathcal{Q}$, depending on whether the readout SNR error $[P(g,1|g)+P(e,0|e)]/2$ is greater than the probability of overlooking a qubit transition.

Finally, we also express the average leakage probability $L=[L_g+L_e]/2$, which we extract using the RILB technique in terms of the readout error channel parameters.
\begin{equation}
\begin{split}
    L_g = P(l_g, 0|g)+P(l_g, 1|g)\\
    L_e = P(l_e, 0|e)+P(l_e, 1|e)
\end{split}
\end{equation}

\section{Readout-induced leakage benchmarking (RILB) technique} 
\label{appendix:rdt_channel_capacity}  
\begin{figure*}[t!]
\centering
\includegraphics[width = 0.95\textwidth]{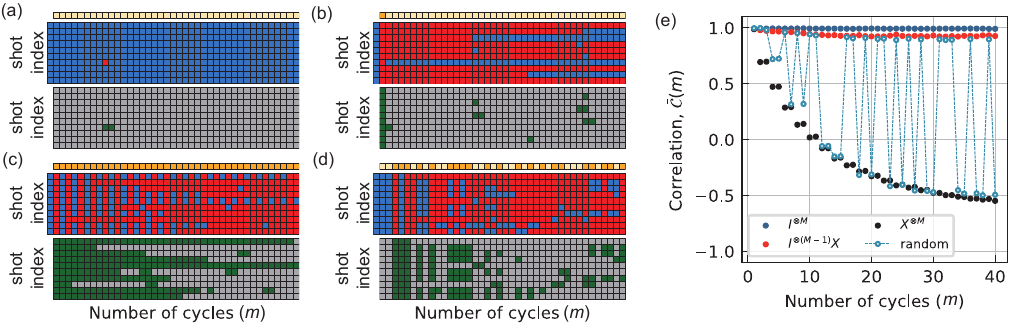}
\caption
{Example of RILB data decoding for a readout pulse corresponding to $\bar{n}_r = 7.6$ for different inputs: (a) $I^{\otimes M}$, (b) $I^{\otimes (M-1)}X$, (c) $X^{\otimes M}$ and (d) a randomized sequence. Top row: input sequence $\{i_m\}$ (orange for $X$, off-white for $I$). Middle panel: 10 example sequences of the readout outcomes (blue for $0$, red for $1$). Bottom panel: corresponding output bit-string $\{o_m\}$ (green for ``flipped", grey for ``not-flipped"). (e) Bit-wise correlation between $\{o_m\}$ and $\{i_m\}$ averaged over $9000$ shots for the four example inputs. 
}
\label{fig:pseudo_syndrome_decoding} 
\end{figure*}

The RILB experiment consists of the following steps. 
\begin{enumerate}
    \item Choose the sequence length $M$ that indicates the total number of cycles. In our experiment, each random sequence of the RILB experiment consists of $M=40$ readout operations after 
    a pre-selection readout that prepares the qubit in $|g\rangle$ with fidelity above $99.9 \%$.
    \item Construct $K$ different randomizations for the input bit-string and generate the corresponding sequences $\mathcal{R}_0\circ\left(\prod_{m = 1}^{M} \mathcal{P}_m \circ \mathcal{R}_m\right)$, where $\mathcal{P}_m \in \{ X, I \}$ is determined by the input bit-string $\{ i_m\}$, and $\mathcal{R}_m$ is the $m$-th readout operation.
    \item From the neighboring pairs of readout outcomes $r_m$ and $r_{m-1}$, construct the output ``flipped-or-not'' bit-string $\{ o_m\}$ such  that $o_m \equiv r_{m-1}\oplus r_m$, where $\oplus$ represents the XOR operation on the two bits. Ideally, it should be identical to the input ``X-or-I'' bit-string.
    \item Compute the bit-wise correlation, $\mathcal{C}_m = 1-2(i_m\oplus o_m)$ for each experimental realization of the sequence, where $\mathcal{C}_m \in \{-1, +1\}$.
    \item Perform $N$ experimental realizations to compute the average bit-wise correlation, $\bar{\mathcal{C}}_m$ for each random sequence. In our experiments, we keep $9000$ post-selected shots after discarding the shots with erroneous state preparation.
    \item Average over different randomizations to obtain the mean success probability $\langle \bar{\mathcal{C}} \rangle_m$. In the data presented here, we have averaged over 98 random input strings. 
\end{enumerate}
In the following, we provide details of the data processing, and provide a theoretical model to extract the leakage rate from the experimental data. 

\subsection{Output bit-string processing}
To illustrate our protocol, we use an example and show the readout outcomes for the calibrated pulse at $\bar{n}_r = 7.6$ for three special input bit-strings: $I^{\otimes M}$ ,  $I^{\otimes (M-1)} X$ and $X^{\otimes M}$ in Fig.~\ref{fig:pseudo_syndrome_decoding}(a-c) (red and blue). We also show the inferred output bit-string (dark green and gray). Ideally, they correspond to $40$ readouts on the $|g\rangle$ states, $40$ readouts on the $|e\rangle$ states, and $40$ readouts with the qubit flipped between each pair of readouts. Finally we show an example of a random input sequence from the RILB experiment in Fig.~\ref{fig:pseudo_syndrome_decoding}(d). 
From the readout outcomes and the inferred output bit-strings in Fig.~\ref{fig:pseudo_syndrome_decoding}(a-d), we observe three distinct types of errors affecting the readout: 
\begin{enumerate}
    \item A discrimination error (due to finite SNR) corrupts one individual readout outcome, leading to two neighboring output bits to be anti-correlated to the input bits. 
    \item A Pauli error (a $|g\rangle\rightarrow|e\rangle$ heating or an $|e\rangle\rightarrow|g\rangle$ decay) still keeps the qubit in the computational subspace,  leading to only one output bit to be anti-correlated. 
    \item A leakage error leaves the bit-flip operations ineffective, making all the following output bits corrupted. 
\end{enumerate}
Only the third kind of error accumulates and results in an exponential decay in the correlation of the input and the output bit-strings.

When the qubit remains in the leakage states, the readout outcomes depend on how the states it leaks into are positioned in the IQ plane. 
We use $P(x|l)$ to represent the probability of the readout outcome being $x$ when the qubit is in the final mixed states in the leakage subspace $l$. 

In the limiting case of $P(1|l) = 1$ (or $P(0|l) = 1$), as the qubit remains in the leakage states the readout outcomes remain fixed at `$1$' (or ``0'') irrespective of the input operations. Thus the output bit will appear anti-correlated when the input operation is `$X$' and correlated when the input operation is `$I$'. When a random input bit string is chosen, the bit-wise correlation $\bar{\mathcal{C}}_m$ (averaged over $N$ experimental realizations) either is close to 1 (when $i_m = I$), or decays exponentially with $m$ (when $i_m = X$). 
In the other limiting case of $P(0|l) = P(1|l) = 1/2$, when the qubit is in the leakage states the readout outcomes are fully randomized. Thus, irrespective of the input operations ($i_m = X$ or $I$) $\bar{\mathcal{C}}_m$ will follow the same exponential decay curves. 

\begin{figure}
\centering
\includegraphics{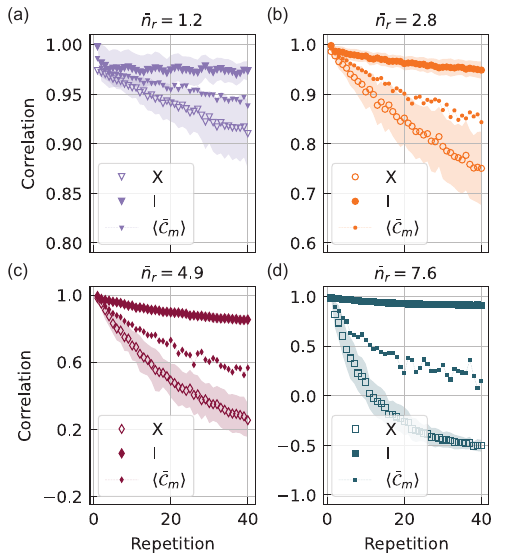}
\caption
{Decay of correlation. Additional fluctuation in the mean correlation is due to finite sampling and the deviation of $p_{\mathbb{X}}$ from 0.5.
}
\label{fig:pseudo_syndrome} 
\end{figure}

In our experiment $P(1|l) \sim 1$, that results in $\bar{\mathcal{C}}_m$ sampling between two distributions following the $I^{\otimes (M-1)}X$ and $X^{\otimes M}$ curves dependent on the input operations, as shown in Fig.~\ref{fig:pseudo_syndrome_decoding}(e). We apply $98$ different randomizations of input sequence $\{i_m\}$ to obtain the mean success probability $\langle \bar{\mathcal{C}}_m \rangle$, plotted in Fig.~\ref{fig:pseudo_syndrome}(a-d) for four different readout powers. 
In addition, we also post-select the shots where $i_m = I$ (or $i_m = X$) and plot the mean correlation for these two classes. The difference between them is representative of the imbalance between $P(1|l)$ and $P(0|l)$. 
The error bar of $\langle \bar{\mathcal{C}}_m \rangle$ is the average of the standard deviation of the two distributions.

\subsection{Leakage model}

We assume a simple leakage model to fit the experimental data in the RILB experiment and estimate a rate for the readout-induced leakage. We consider a larger Hilbert space consisting of the $d_1 = 2$-dimensional computational subspace $\mathcal{X}_1$, and a $d_2$-dimensional leakage subspace $\mathcal{X}_2$, with $\mathbb{1}_1$ and $\mathbb{1}_2$ being the projectors onto the respective subspaces~\cite{wood_lrb_2018}. Thus the full dynamics happen in a $(d_1+d_2)$-dimensional direct-sum state space $\mathcal{X}_1\oplus\mathcal{X}_2$. Let $\rho$ be an arbitrary state defined in the $(d_1+d_2)$-dimensional Hilbert space. The total leakage population of state $\rho$ is then defined as $\mathfrak{L}(\rho) = {\rm{Tr}}[\mathbb{1}_2\rho]$. We can describe the leakage error channel for a single cycle as a completely positive trace preserving (CPTP) map, $\mathcal{E}_L$ in this $(d_1+d_2)$-dimensional space. Therefore, if we initiate the system in a given state $|\psi\rangle$ within the computational subspace, the leakage rate for one cycle is given by: $p_l = \mathfrak{L}\left(\mathcal{E}_L\left[|\psi_1\rangle\langle\psi_1|\right]\right)$. However, instead of focusing on 
the leakage rate from a particular state, we are rather interested in the average leakage rate $L_\uparrow$ of the channel $\mathcal{E}_L$ when acted on a random input state in the computational subspace:
\begin{equation}
    L_\uparrow\left(\mathcal{E}_L\right) = \int d\psi_1\mathfrak{L}\left(\mathcal{E}_L\left[|\psi_1\rangle\langle\psi_1|\right]\right)= \mathfrak{L}\left(\mathcal{E}_L\left(\frac{\mathbb{1}_1}{d_1}\right)\right)
\label{eq:dle_avg}
\end{equation}
Note that we can still put a bound on the worst case leakage from any individual state $\rho'$ in the computational subspace (assuming all the leakage happens from that state):
\begin{equation}
    \mathfrak{L}_{\rm{max}}\left(\mathcal{E}_L[\rho_0]\right) \leq d_1 L_\uparrow (\mathcal{E}_L).
\end{equation}
In the same spirit, we can also define an average \textit{seepage} rate,  the rate at which the population may decay back into the computational subspace, caused by the error channel by averaging over all leakage states $|\psi_l\rangle$:
\begin{equation}
    L_\downarrow\left(\mathcal{E}_L\right) = 1-\int d\psi_l\mathfrak{L}\left(\mathcal{E}_L\left[|\psi_l\rangle\langle\psi_l|\right]\right)=1-\mathfrak{L}\left(\mathcal{E}_L\left(\frac{\mathbb{1}_2}{d_2}\right)\right),
\label{eq:dle_avg2}
\end{equation}

To describe the leakage process, we assume that there is no coherent exchange between the computation states and the leakage states. We define a completely depolarizing channel $\mathcal{D}_{ij}(\rho)$ between subspaces $\mathcal{X}_i$ and $\mathcal{X}_j$:
\begin{equation}
    \mathcal{D}_{ij}(\rho) = {\rm{Tr}}\left[\mathbb{1}_j\rho\right]\frac{\mathbb{1}_i}{d_i}, \quad i, j \in \{1, 2\}
\end{equation}
$\mathcal{D}_{12}$ and $\mathcal{D}_{21}$ destroys any coherence in the leakage and seepage process and $\mathcal{D}_{22}$ destroys any coherent dynamics in the leakage subspace. This is an important \textit{assumption} that eliminates any memory effects to arrive at an exponential decay model.

The leakage error can then be described as a completely positive trace preserving (CPTP) map, $\mathcal{E}_L$:
\begin{equation}
    \mathcal{E}_L = (1-L_\uparrow)\mathcal{E}_1 + L_\uparrow\mathcal{D}_{21} +
    L_\downarrow\mathcal{D}_{12} +
    (1-L_\downarrow)\mathcal{D}_{22},
    \label{eq:dle}
\end{equation}
where $\mathcal{E}_1$, in general, is an arbitrary CPTP map within the computational subspace. The readout operation combined with the randomized bit-flips turns it into a completely depolarizing channel, $\mathcal{D}_{11}$ that may also absorb any Pauli error.

We can compute the leakage population after $m$ applications of the cycle by raising the error channel $\mathcal{E}_L$ to power $m$ :
\begin{equation}
    \mathfrak{L}\left(\mathcal{E}_L^m[\rho]\right) = {\rm{Tr}}\left[\mathbb{1}_2\mathcal{E}_L^m[\rho]\right] = {\rm{Tr}}\left[\rho\left(\mathcal{E}_L^\dagger\right)^m[\mathbb{1}_2]\right]
\label{eq:err_exp}
\end{equation}
To arrive at the last expression, we have used the properties of matrix adjugates, ${\rm{Tr}}[A] = {\rm{Tr}}[A^\dagger]$ and $(AB)^\dagger = B^\dagger A^\dagger$.

As the operators $\mathcal{D}_{ij}$ are all mutually orthogonal, we can express the superoperator for $\mathcal{E}_L$ with respect to the bases, $||\mathbb{1}_1\rangle\rangle \equiv (1, 0)$ and $||\mathbb{1}_2\rangle\rangle  \equiv (0, 1)$ by a $2\times2$ matrix:

\begin{equation}
    \mathcal{S}_{\mathcal{E}_L}=
    \begin{pmatrix}
        1-L_\uparrow & L_\downarrow\\
        L_\uparrow & 1-L_\downarrow
    \end{pmatrix}, 
\label{eq:leakage_model}
\end{equation}

The superopreator for the Hermitian conjugate matrix (${\mathcal{E}_L^\dagger}$) and its $m$-th power can be written as:
\begin{equation}
\begin{split}
    \mathcal{S}_{\mathcal{E}_L^\dagger}=&
    \begin{pmatrix}
        1-L_\uparrow & L_\uparrow\\
        L_\downarrow & 1-L_\downarrow
    \end{pmatrix},\\
       \mathcal{S}_{\mathcal{E}_L^\dagger}^m=&\frac{1}{L}
    \begin{pmatrix}
        L_\downarrow & 
        L_\uparrow \\
        L_\downarrow  & 
        L_\uparrow
    \end{pmatrix} +
    \frac{(1-L)^m}{L}
    \begin{pmatrix}
        L_\uparrow & 
        -L_\uparrow \\
        -L_\downarrow  & 
        L_\downarrow
    \end{pmatrix}
\end{split}
\label{eq:leakage_model_conjugate}
\end{equation} 
We can apply this superoperator on the basis state $||\mathbb{1}_2\rangle\rangle$ to obtain:
\begin{equation}
\begin{split} \mathcal{S}_{\mathcal{E}_L^\dagger}^m||\mathbb{1}_2\rangle\rangle &= 
   \frac{L_\uparrow}{L}
    (||\mathbb{1}_1\rangle\rangle+||\mathbb{1}_2\rangle\rangle) \\
    &-
    \frac{(1-L)^m}{L}
    (L_\uparrow||\mathbb{1}_1\rangle\rangle-L_\downarrow||\mathbb{1}_2\rangle\rangle)
\end{split}
\label{eq:sup_op_exp}
\end{equation}
Hence, using Eq.~\ref{eq:err_exp} and Eq.~\ref{eq:sup_op_exp} we find the expression the total leakage population after $m$ application of the detection cycle when start from an initial state with a leakage population $p_{\rm{ini}}$:
\begin{equation}
\begin{split}
    \mathfrak{L}\left(\mathcal{E}_L^m[\rho]\right) &= \Bigl((1-p_{\rm{ini}})\langle\langle\mathbb{1}_1||+p_{\rm{ini}}\langle\langle\mathbb{1}_2||\Bigr)\mathcal{S}_{\mathcal{E}_L^\dagger}^m||\mathbb{1}_2\rangle\rangle\\
    & = \frac{L_\uparrow}{L}-(1-L)^m\left(\frac{L_\uparrow}{L}-p_{\rm{ini}}\right)   
\end{split}
\end{equation}
If the \textit{bit-flip operation is perfect} and either $P(0|\psi_l)=0$
or $P(1|\psi_l)=0$ and there is \textit{no other readout error apart from leakage}, the average bit-wise correlation between the input string and the output string is given by:
\begin{equation}
\begin{split}
    \langle\bar{\mathcal{C}}_m\rangle &= \frac{(+1)}{2} +\frac{(+1)\left[1-\mathfrak{L}\left(\mathcal{E}_L^m[\rho]\right)\right]+(-1)\left[\mathfrak{L}\left(\mathcal{E}_L^m[\rho]\right)\right]}{2}\\
    &= 1-  \mathfrak{L}\left(\mathcal{E}_L^m[\rho]\right) = \frac{L_\downarrow}{L}-(1-L)^m\left(\frac{L_\uparrow}{L}-p_{\rm{ini}}\right),
\end{split}
\end{equation}
And we can individually estimate $L_\uparrow$, $L_\downarrow$ and $p_{\rm{ini}}$. However, in presence of other errors such as, imperfect bit-flip operations ($\epsilon_{\pi}$), Pauli errors ($\epsilon_{\Gamma}$), readout discrimination error ($\epsilon_{\rm{SNR}}$), and when $0<P(0|\psi_l)<1$, the prefactors get modified and they can be combined into two constants to be determined from the fit:
\begin{equation}
    \langle\bar{\mathcal{C}}_m\rangle=\frac{1}{2}\left(A+ B(1-L)^m\right),
\end{equation}
We use this expression to fit the experimental data and estimate the combined leakage-seepage rate $L$ for the four readout pulses optimized for different powers.